\documentclass[10pt]{article}
\usepackage{setspace}
\usepackage{xspace}
\date{}
\usepackage{algorithm}
\usepackage{algpseudocode}
\usepackage{wrapfig}
\usepackage[margin=1in]{geometry}
\usepackage{graphicx}
\usepackage{color}
\usepackage{cite}
\usepackage{url}
\usepackage{appendix}
\graphicspath{{figures/}}

\usepackage{times}
\usepackage[small,compact]{titlesec}
\usepackage[small,it]{caption}
\usepackage{setspace}
  \renewenvironment{thebibliography}[1]{%
    \begin{oldthebibliography}{#1}%
      \setlength{\parskip}{0.3ex}%
      \setlength{\itemsep}{0ex}%
  }%
  {%
    \end{oldthebibliography}%
  }

\begin{document}

\begin{singlespace}
\title{Modern Gyrokinetic Particle-In-Cell Simulation of Fusion Plasmas on Top Supercomputers} 

%\author{authors - omitted for blind review}
%\vspace{-1.5in}
\author{ \small
Bei Wang$^1$ , Stephane Ethier$^2$, William Tang$^{1,2}$, Khaled Z. Ibrahim$^3$, Kamesh Madduri$^4$, Samuel Williams$^3$, Leonid Oliker$^3$\\
{\em \small $^3$ Princeton Institute of Computational Science and Engineering, Princeton University, Princeton, NJ, USA}\\
{\em \small $^4$ Princeton Plasma Physics Laboratory, Princeton, NJ, USA}\\
{\em \small $^1$ CRD, Lawrence Berkeley National Laboratory, Berkeley, USA}\\
{\em \small $^2$ The Pennsylvania State University, University Park, PA, USA}\\
{\em  \small beiwang@princeton.edu, \{\ ethier, tang\ \}@pppl.gov, \{\ KZIbrahim, SWWilliams, LOliker\ \}@lbl.gov,  Madduri@cse.psu.edu}\\
%\vspace{-.3in}
}

\maketitle

%\begin{abstract}
Abstract -- The Gyrokinetic Toroidal Code at Princeton (GTC-P) is a highly scalable and portable particle-in-cell (PIC) code.  It solves the 5D Vlasov-Poisson equation featuring efficient utilization of modern parallel computer architectures at the 
petascale and beyond. Motivated by the goal of developing a modern code capable of dealing with the physics challenge of increasing problem size with sufficient resolution, new thread-level optimizations have been introduced as well as a key additional domain decomposition. 
GTC-P's multiple levels of parallelism, including inter-node 2D domain decomposition and particle decomposition, as well as intra-node shared memory partition and vectorization have enabled pushing the scalability of the 
PIC method to extreme computational scales. In this paper, we describe the methods developed to build a highly parallelized PIC code across a broad range of supercomputer designs.  
This particularly includes implementations on heterogeneous systems using NVIDIA GPU accelerators and Intel Xeon Phi (MIC) co-processors and performance comparisons with state-of-the-art homogeneous HPC systems such as Blue Gene/Q. 
New discovery science capabilities in the magnetic fusion energy application domain are enabled, including investigations of Ion-Temperature-Gradient (ITG) driven turbulence simulations with unprecedented 
spatial resolution and long temporal duration. Performance studies with realistic fusion experimental parameters are carried out on multiple supercomputing systems spanning a wide range of cache capacities, cache-sharing configurations, memory bandwidth, interconnects and network topologies. 
These performance comparisons using a realistic discovery-science-capable domain application code provide valuable insights on optimization techniques across one of the broadest sets of current high-end computing platforms worldwide.
%\end{abstract}
\end{singlespace}

\section{Introduction}
The Gyrokinetic Toroidal Code (GTC) \cite{LHL98} is a particle-in-cell (PIC) code that solves the five-dimensional (5D) gyrokinetic Vlasov equation in full, global torus geometry to address kinetic turbulence issues in magnetically-confined fusion experimental facilities known as tokamaks \cite{Tang09}.
This code was developed using FORTRAN 90 and features three levels of parallelism: a one-dimensional domain decomposition in the toroidal dimension, a particle decomposition in each 
toroidal domain, and multithreaded, shared memory loop-level parallelism implemented with OpenMP \cite{ETL05, ETW08}. The method scales well to a large number of processor cores with respect to the number of particles. However, the approach can suffer from performance bottlenecks when 
dealing with significantly increased problem size, e.g., simulating large-scale plasmas, such as the burning plasma experimental facility ITER, the largest device currently under construction in France \cite{ITER}.
This is due to the increasing grid-related computation and memory requirement on each process. With only 
one-dimensional domain decomposition in the toroidal dimension, this algorithm has each process keep a copy of the full poloidal grid, and the number of grid points in the poloidal grid increases 4x for a plasma device of 2x in minor radius. In order to effectively address the open questions in fusion plasma physics, such as the scaling of the energy 
confinement time with system size \cite{Lin02,McMillan10,Idomura2014,Tang14}, a key additional level of domain decomposition in the radial dimension was introduced.
In this new version named GTC-Princeton or GTC-P \cite{EACO10,Adams07}, the additional domain decomposition substantially reduces the memory requirements and thus enables the code to deal with increased problem size, including being
able to scale to ITER-size simulations. 

The illustrative application domain on which the present studies are focused begins with the fact that the fusion of light nuclides forms the basis of energy release in the universe and can potentially 
be harnessed and used as a clean and sustainable supply of energy on Earth. Magnetic confinement fusion devices in which very high temperature plasma is confined in a tokamak, a donut-shaped vacuum vessel, is one of the most promising approaches to fusion reactors.  
In such a system, the interplay between the complex trajectories of individual charged particles and the collective effects arising from the long range nature of electromagnetic forces lead to a wide range of waves, oscillations, and instabilities. 
These include larger-scale (macro) instabilities that produce rapid topological changes in the device resulting in a catastrophic loss of fusion power, as well as smaller-scale (micro) instabilities that gradually leak energy and thus affect performance and 
economic viability. Turbulent fluctuations (\textquotedblleft microturbulence\textquotedblright) are instabilities caused by the unavoidable spatial variations (gradients) in a plasma system. They can significantly increase the transport rate of heat, particles, 
and momentum across the confining magnetic field and severely limit the energy confinement time for a given machine size. High-fidelity predictive numerical experiments based on 5D gyrokinetic simulations play a significant role in understanding, 
predicting and possibly controlling these instabilities. However, gyrokinetic simulations are extremely compute-intensive since the phase space distribution function in five dimensions needs to be accurately resolved.  This is a key motivating factor for developing 
highly-optimized modern PIC codes that can effectively carry out microturbulence simulations capable of being deployed on the most powerful supercomputing systems. 

The basic particle method has long been a well established approach for simulating the behavior of charged particles interacting with each other through pair-wise electromagnetic forces.  
At each time step, the particle properties are updated according to these calculated forces.  For applications on powerful modern supercomputers with deep cache hierarchy, 
a pure particle method is efficient with respect to both locality and arithmetic intensity (compute bound). Unfortunately, the O($N^2$) complexity makes a particle method impractical 
for plasma simulations using millions of particles per process.  Rather than calculating O($N^2$) forces, the particle-in-cell (PIC) method, which was introduced 
by J. Dawson and N. Birdsall \cite{Birdsall91} in 1968, employs a grid as the medium to calculate the long range electromagnetic forces.  
This reduces the complexity from O($N^2$) to O($N+M\log M$), where $M$ is the number of grid points that is generally much smaller than $N$.  
However, achieving high parallel and architectural efficiency is a significant challenge for PIC methods due to potential fine-grained data hazards, irregular data access, and low arithmetic 
intensity.  Attaining performance becomes even more complex as HPC technology evolves towards vast on-node parallelism in modern multi- and many-core designs.  A deep understanding of how to improve the associated scalability will have a wide-ranging influence on numerous physical applications, which 
use particle-mesh algorithms, including molecular dynamics, cosmology, accelerator physics, and plasma physics.  

Driven by chip- and system power limitations, the HPC community is moving to the era of multi- and many-core architectures with greatly increased thread and vector parallelism on shared memory processors. 
While the family of the GTC FORTRAN codes (the original GTC \cite{ETL05} and GTC-P \cite{EACO10,Adams07}) continue to scale well to a large number of processes, the level and efficiency of multithreading need to be increased significantly to take full advantage of the upcoming technology.
In order to benefit from the computer science advances in exploiting multi-threading capabilities to facilitate large-scale simulations on modern systems, we have strived to develop  
a thread- and accelerator-optimized version of the code over the last five years. These efforts commenced with a complete rewrite of the original GTC FORTRAN code in the C language. The need for this change arose from our 2009 study of the GTC charge accumulation step, in which several low-level multi-threaded algorithms were implemented using \emph{Pthread} programming~\cite{MWE09} and compared for performance. The best algorithm was later re-implemented with OpenMP, replacing the much more complicated Pthread constructs. The C version also simplified the process of porting GTC-P to GPU and Intel Xeon Phi.
Novel multi-core centric optimizations were introduced to enhance the performance of the code \cite{MWE09, MII11, Madduri11}. The earlier efforts also included the development of the corresponding GPU version using the CUDA language \cite{Madduri11,Ibrahim13}. 
The journey continued with the development of a highly efficient radial domain decomposition \cite{Wang13}. Specifically, we carefully addressed several issues identified in the original GTC-P FORTRAN code, such as load imbalance and the lack of multithreading
capability for some grid-based subroutines. The multiple levels of parallelism, including inter-node 2D domain decomposition and particle decomposition, and intra-node shared memory partition, as well as vectorization within each core, have allowed the new GTC-P code to efficiently scale to the full capability of several 
top computer systems, including Sequoia at Lawrence Livermore National Laboratory (US) and Mira at the Argonne Leadership Computing Facility (US) \cite{Wang13}. 

In this paper, we describe our approach to developing such a highly parallel PIC code that has the unique capability of efficiently simulating large size plasmas on both homogeneous and heterogeneous systems.
In particular, we discuss our recent experience in porting and optimizing GTC-P on NVIDIA GPU- and Intel Xeon Phi- (MIC) accelerated systems. The performance studies with production simulation parameters are performed on numerous supercomputing platforms including Titan at the Oak Ridge Leadership Computing Facility (US), Mira at the Argonne Leadership Computing Facility (US), Piz Daint at the Swiss National Supercomputing Center (Switzerland), Edison at the National Energy Research Scientific Computing Center (US), 
Stampede at the Texas Advanced Computing Center (US) and Blue Waters at the National Center for Supercomputing Applications (US). This is the first study that examines the production code with \textquotedblleft true weak scaling study\textquotedblright\ on both large scale homogeneous and heterogeneous HPC systems, as opposed to 
our earlier research that focused on either large size (ITER-size) simulations on CPU systems \cite{Wang13} or small to moderate size simulations on GPU systems \cite{Madduri11,Ibrahim13}.
Given that the PIC algorithm consists of several phases of varying computational intensity, memory access patterns, and communication patterns, our study enables exploration with respect to computer science contributions as well as from a full-scale discovery-science-capable applications standpoint.  
With regard to the important challenge of dealing with code portability on advanced supercomputing platforms, another contribution from these studies involves providing a quantitative comparison of the number of \textquotedblleft lines of code changes\textquotedblright~and associated speedup obtained from our experience in developing 
the modern GTC-P code across a wide range of architectures.

The sections below are organized as follows. In \S \ref{sec:equations} we first review the 5D gyrokinetic Vlasov-Poisson equation used for studying low-frequency microturbulence in fusion plasmas. 
Then we describe the numerical discretization for the system based on the PIC method. In \S \ref{sec:GTCP}, the modern GTC-P code is introduced. We discuss its main kernels as well as the parallelization and 
optimization deployed. In particular, this paper provides results from our new optimization techniques for GPUs and description of our experience in porting and optimizing the code on the Intel Xeon Phi system.
The performance analysis and results of high resolution, long time simulations of Ion Temperature Gradient (ITG) driven turbulence are given at the end of \S \ref{sec:results}.

\section{Gyrokinetic Vlasov Poisson System}\label{sec:equations}

\subsection{Governing Equations}
The study of low-frequency microturbulence in high-temperature, magnetically confined plasmas requires the use of the kinetic model described by the Vlasov equation in six-dimensional phase space. 
In the gyrokinetic approach~\cite{Lee87}, the dynamics of the high frequency cyclotron motion of the charged particles in the strong magnetic field is averaged out, reducing the six-dimensional equation
to a five-dimensional gyrokinetic equation:
\begin{equation}
 \frac{df_\alpha}{dt} = \frac{\partial f_\alpha}{\partial t} + \frac{d\mathbf{R}}{dt} \cdot \frac{\partial f_\alpha}{\partial \mathbf{R}} +
\frac{d v_{\|}}{dt}\frac{\partial f_\alpha}{\partial v_{\|}} = 0,
\label{eqn:f}
\end{equation}
\begin{equation}
 \frac{d \mathbf{R}}{dt} = v_{\|}\hat{\mathbf{b}} + \mathbf{v}_{E} + \mathbf{v}_d,
\label{eqn:dRdt}
\end{equation}
\begin{equation}
 \frac{d v_{\|}}{dt} = -\frac{1}{m_\alpha}\mathbf{b}^\ast\cdot(\mu \nabla B + Z_\alpha \nabla \bar{\phi})
\label{eqn:dvdt}
\end{equation}
\begin{equation}
 \mathbf{b}^\ast  = \hat{\mathbf{b}} + \frac{v_{\|}}{\Omega_\alpha} \nabla \times \hat{\mathbf{b}}
  \approx  \hat{\mathbf{b}} + \frac{v_{\|}}{\Omega_\alpha} \frac{\mathbf{\hat{b}} \times \nabla B}{B}, 
\end{equation}
where $f_\alpha(\mathbf{R}, v_{\|}, \mu)$ is the five-dimensional phase space distribution function of species $\alpha$ in the gyrocenter 
coordinates $\mathbf{R}$. $Z_\alpha$ is the particle charge. $m_\alpha$ is the particle mass. $\mathbf{B} = B \hat{\mathbf{b}}$ is the nonuniform 
equilibrium magnetic field. $\mu = \frac{m_\alpha v_{\bot}^2}{2B}$ is the magnetic moment.  $\bar{\phi}$ is the gyrophase-averaged potential.
$\mathbf{v}_E$, $\mathbf{v}_c$ and $\mathbf{v}_g$ are the $E \times B$ drift, magnetic curvature drift, and grad-B drift velocities, which take the forms
\begin{equation}
 \mathbf{v}_E = \frac{ c \mathbf{\hat{b}} \times \nabla \bar{\phi}}{B},
\end{equation}
\begin{equation}
 \mathbf{v}_c = \frac{v_{\|}^2}{\Omega_\alpha} \nabla \times \hat{\mathbf{b}} \approx \frac{v_{\|}^2}{\Omega_\alpha} \frac{\mathbf{\hat{b}} \times \nabla B}{B},
\end{equation}
\begin{equation}
 \mathbf{v}_g = \frac{1}{m_\alpha \Omega_\alpha} \hat{\mathbf{b}} \times \mu \nabla {B},
\end{equation}
\begin{equation}
 \mathbf{v}_d = \mathbf{v}_c + \mathbf{v}_g = (\frac{v_{\|}^2}{\Omega_\alpha} + \frac{c\mu}{Z_\alpha}) \frac{\hat{\mathbf{b}} \times \nabla {B}}{B}.
\end{equation}
where $c$ is the speed of the light and $\Omega_\alpha=\frac{Z_\alpha B}{ m_\alpha c}$ is the gyrofrequency. We use the identity
\begin{equation}
 \nabla \times \mathbf{B} = B \nabla \times \hat{\mathbf{b}} + \nabla B \times \hat{\mathbf{b}} = 0
\end{equation}
in the low $\beta$ limit.

Consider a simple plasma of electrons and ions, the corresponding gyrokinetic Poisson's equation is
\begin{equation}
\nabla^2\phi - \frac{4\pi Z_i^2 n_{i0}}{T_i}(\phi - \tilde{\phi}) = -4 \pi (Z_i \bar{n}_i - e n_e),
\label{eqn:gyro_poisson}
\end{equation}
where the second gyrophase-averaged potential is 
\begin{equation}
 \tilde{\phi} = \frac{1}{2 \pi} \int \bar{\phi}(\mathbf{R})F_{Mi}(\mathbf{R}, v_{\|}, \mu)
\delta(\mathbf{R-x-\rho})d\mathbf{R}d\mu dv_{\|} d\varphi,
\label{eqn:tildephi}
\end{equation}
and 
\begin{equation}
 \bar{\phi} = \frac{1}{2 \pi}\int \phi(\mathbf{x}) \delta(\mathbf{R-x-\rho})d\mathbf{x}d\mu dv_{\|} d\varphi.
\end{equation}
$F_{Mi}(\mathbf{R}, v_{\|}, \mu)$ is assumed to be Maxwellian. $\bar{n}_i$ is the gyrophase-averaged charge density
\begin{equation}
 \bar{n}_i = \frac{1}{2 \pi}\int f_i(\mathbf{R}) \delta(\mathbf{R-x-\rho})d\mathbf{x}d\mu dv_{\|} d\varphi.
\end{equation}
Equation \ref{eqn:gyro_poisson} can be interpreted as a normal Poisson equation plus an ion polarization term (the second term on the left). In 
the gyrokinetic ordering limit, we usually drop the three dimensional Debye shielding term (the first term) in the gyrokinetic Poisson's equation as the Debye shielding term is much smaller than the ion polarization
term.

 In the case of adiabatic electron where $\frac{\delta n_e}{n_{e0}} = \frac{e \phi}{T_e}$, the 
Poisson equation is reduced to
 \begin{equation}
\frac{ Z_i^2 n_{i0}}{T_i}(\phi - \tilde{\phi}) + \frac{e^2n_{e0}}{T_e}\phi = Z_i \bar{\delta n}_i,
\label{eqn:poisson}
 \end{equation}
where $n_{i0}$ and $n_{e0}$ are ion and electron densities at equilibrium.

\subsection{Numerical Discretization}
In the PIC method, the five dimensional distribution function is represented by a set of particles. Specifically, the PIC simulations are being carried out using marker particles representing a small volume of phase space characterized by position, velocity, and weight.  
After being initialized, the particles follow trajectories computed from the characteristic curves given by the Vlasov equation. There are usually two approaches for the initial loading of the particles. They can be loaded 
uniformly in phase space and assigned different weights that are proportional to the value of the distribution function, or, alternatively, they can be loaded following the distribution function where
each particle carries the same weight. The latter is based on the importance sampling approach of the Monte Carlo method~\cite{Aydemir94}, and thus is preferred as it produces a low-noise loading. To avoid particle 
load imbalance due to spatial gradient in the distribution function, we use a uniform background density function based on multi-scale expansion \cite{Lee2011}.

In order to address the well-known issue of particle noise in PIC methods, we use the so-called $\delta f$ method \cite{Parker93}.
In our simulations, the particles discretize only the perturbation $\delta f$ with respect
to an equilibrium state $f_0$. The $\delta f$ method reduces noise by replacing as much of the Monte Carlo estimate as possible by an analytical formula \cite{Parker93, Aydemir94} describing $f_0$, and hence reducing the variance . 
However, it does not control the noise increase during the long time particle evolution. For carrying out long time simulations, we introduce coarse-graining to the system, also
called phase space remapping~\cite{Chen07,Wang11,Wang12}. The phase space grid works as a natural numerical dissipation model to stabilize the collisionless system in long time simulations.
The system of equations are advanced using numerical integration, e.g., the second order Runge Kutta method.

\begin{figure}
\centering
\includegraphics[width=0.4\columnwidth]{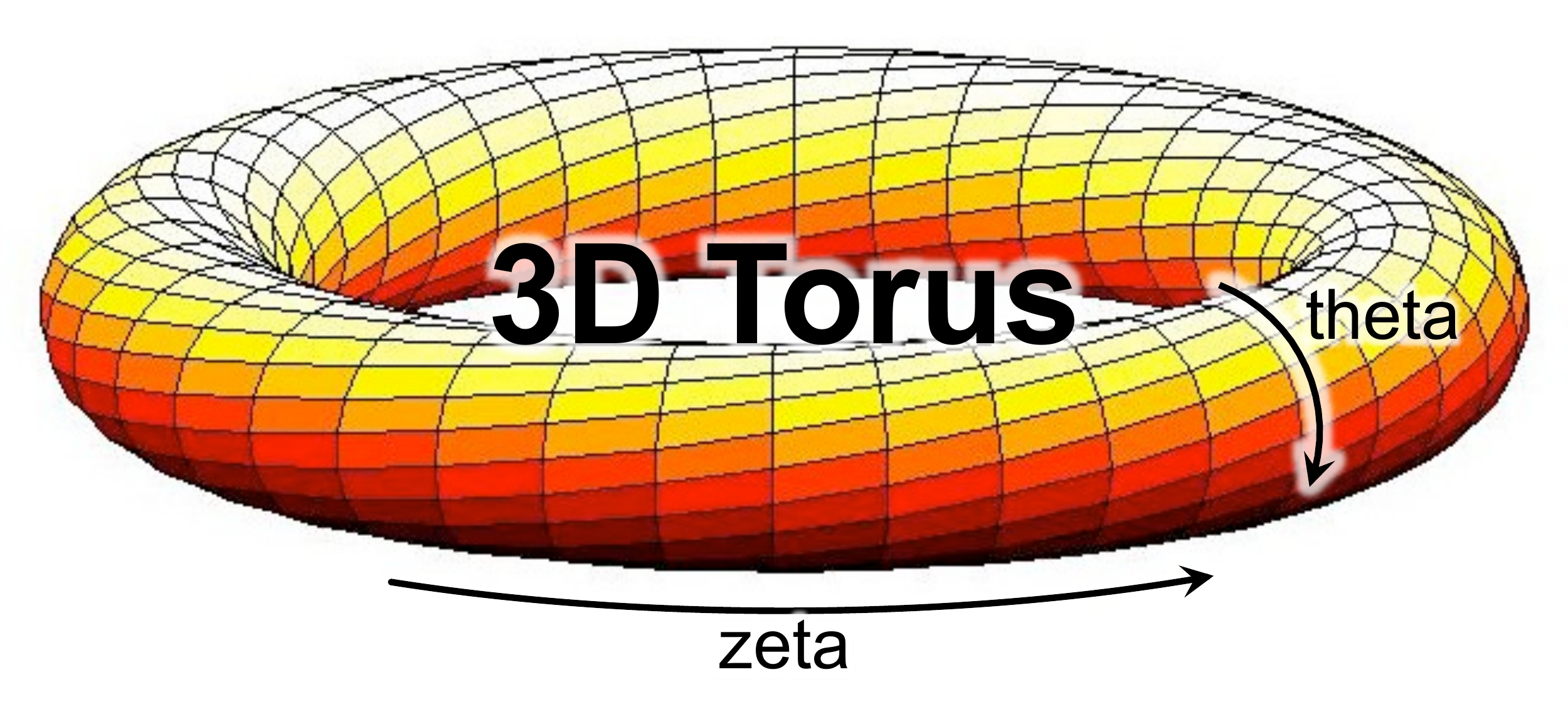}
\includegraphics[width=0.3\columnwidth]{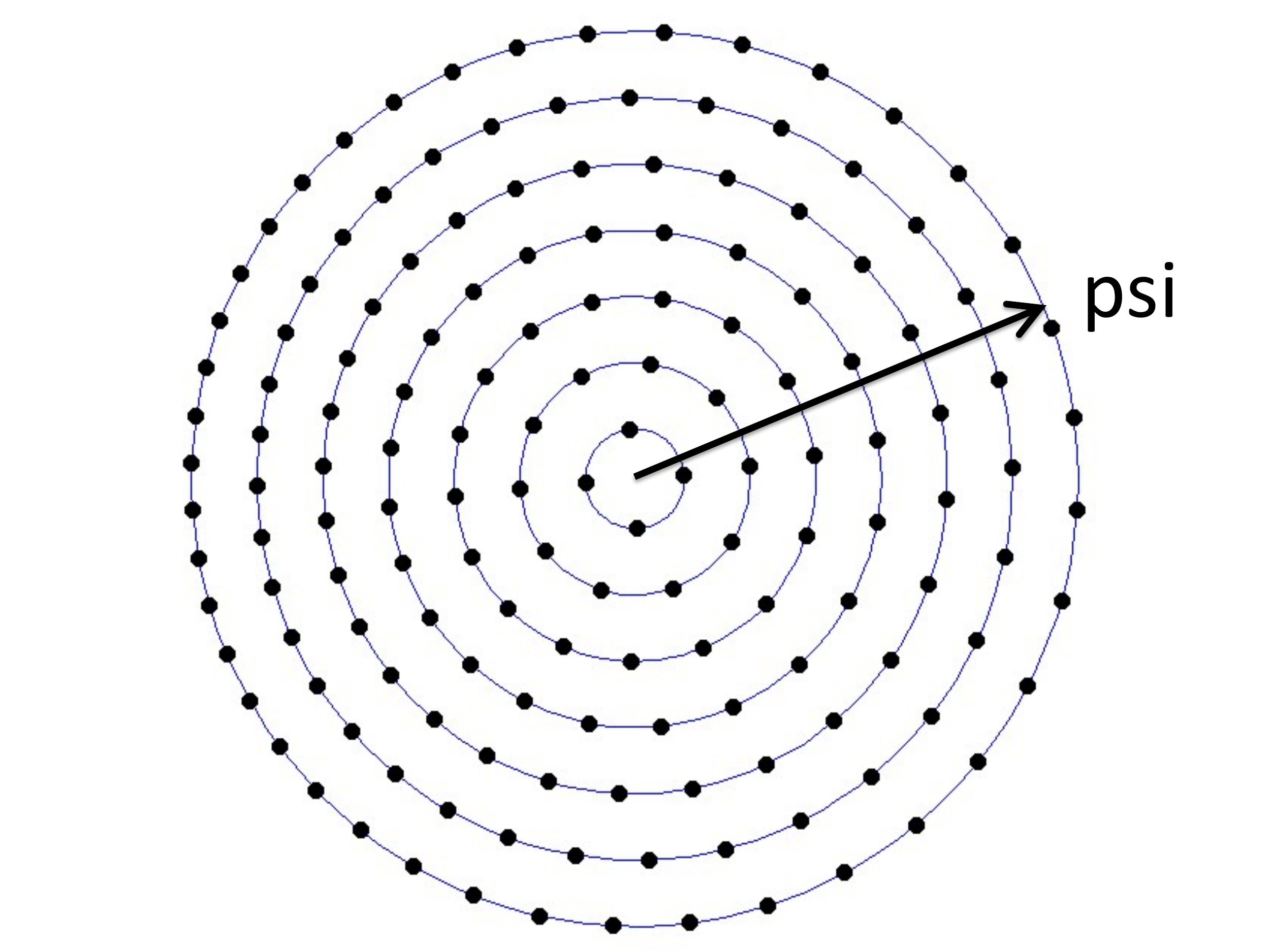}
\caption{\small An illustration of 3D toroidal grid.}
\label{fig:GTC}

\end{figure}

In the PIC method, the Poisson equation is solved on a grid. In GTC-P, we utilize a highly specialized grid that follows the magnetic field lines as they twist around the torus due to the magnetic field pitch~(Figure~\ref{fig:GTC}). This allows the code to take advantage of the strong anisotropy between the fast motion of the particles along the field lines and their slow motion across them. With such a grid, the same accuracy can be achieved by using much fewer poloidal planes ($1/100$) than a non field-aligned grid.
Thus, a production simulation generally consists of only 32 or 64 high resolution poloidal planes wrapped around the torus. 
The 3D torus geometry is first discretized uniformly in the toroidal dimension. Each 2D poloidal plane (cross section of the torus) is represented by an unstructured grid with uniform spacing in the radial 
direction (\emph{psi}) and approximate uniform arc length spacing in the poloidal direction (\emph{theta}). To follow the magnetic field line, each concentric circle forming the radial grid of a poloidal plane shifts by a specific poloidal angle as one goes around the torus (see Figure~\ref{fig:GTC}).
As discussed earlier, the gyrokinetic approximation neglects
the three dimensional Debye shielding term in the Poisson equation, which reduces the complex 3D grid-based solve to a fixed number of independent 2D solves.
On each 2D plane, the four points algorithm \cite{Lin95} is used to approximate the second order gyro-averaged operator (Equation \ref{eqn:tildephi}) on the unstructured grid.
The resulting diagonally dominant matrix is then solved by a weighted Jacobi iterative solver \cite{Lin95}.

\section{The modern GTC-P code}\label{sec:GTCP}
GTC-P is developed under the philosophy that in order to leverage the computer science and applied mathematics advances in a timely matter, the physical model needs to be sufficiently simple but nevertheless complete 
for studying the scaling of turbulent transport spanning the range from present generation experiments to the large ITER-scale plasmas. 
This leads to the fact that GTC-P includes the model that takes into account all the toroidicity effects, such as the curvature drift and multiple rational surfaces, but not the 
non-circular cross-section effects or the fully electromagnetic, non-adiabatic electron dynamics found in some of the other gyrokinetic codes at the current stage~\cite{GTS,ORB5,GYRO,Ku09}.
These techniques reduce the complexity of developing the algorithmic advances required to take advantage of rapidly evolving architectural platforms.
It is worth mentioning that the modern GTC-P code shares the same physical model as the original FORTRAN code that has a long history of producing important scientific discoveries via state-of-art computers \cite{LHL98,Lin02}. 
Besides the implementation differences, the major difference is the dissipation models used in these two codes. While the original FORTRAN version used an artificial heat bath model,
the modern GTC-P introduced several numerical dissipation models \cite{Chen07,Wang11,Mcmillan08} to stabilize long time simulation. 

\subsection{Main kernels}
Each time step of GTC-P requires the execution of six principal computational kernels.  The kernels present a number of intra- and inter-node performance and scalability challenges when running on today's multicore and accelerated supercomputers.  We highlight the kernels and their challenges here.

\begin{figure}
\begin{center}
\vspace{-15pt}
\includegraphics[width=.3\textwidth]{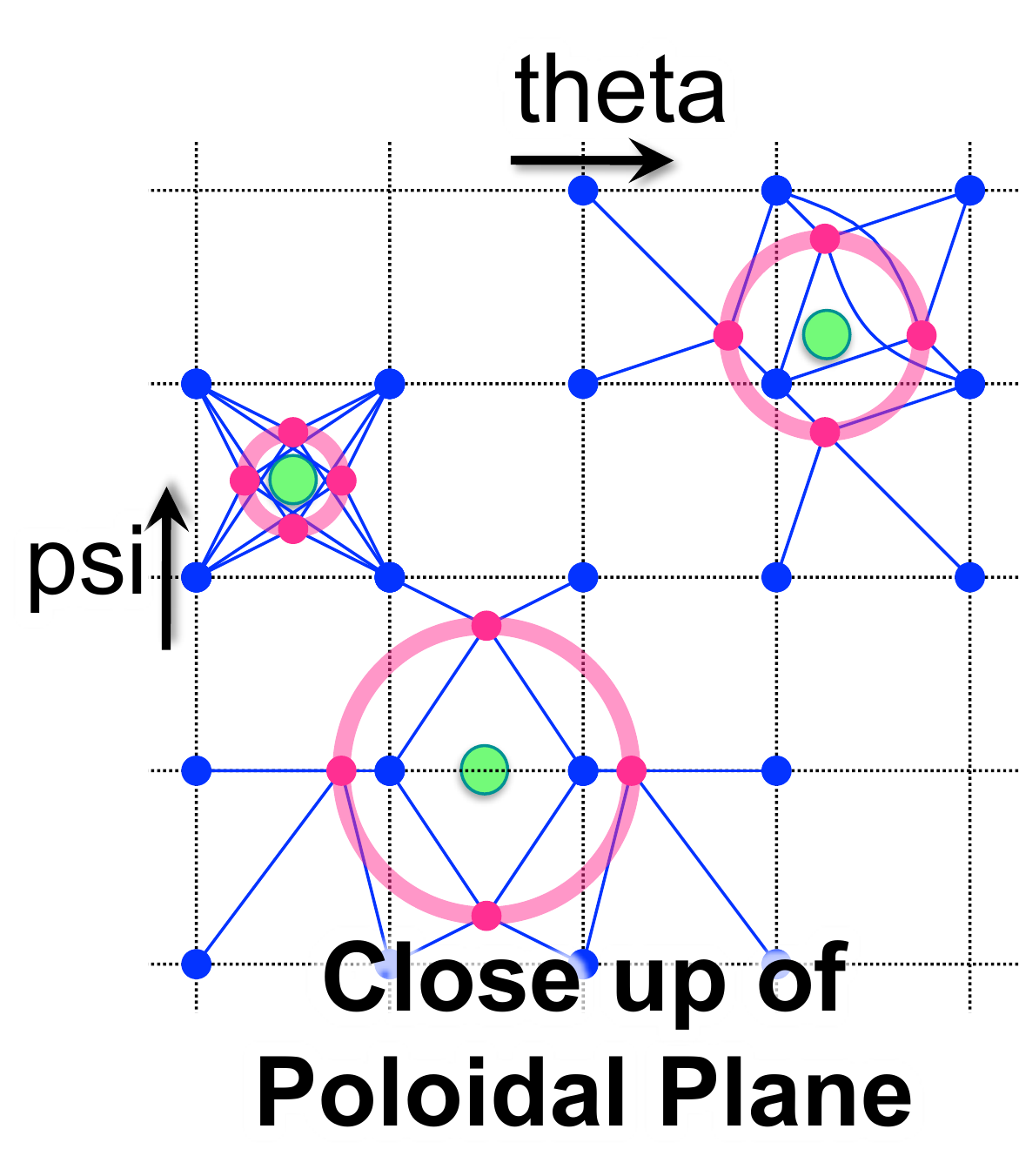}
\vspace{-15pt}
\caption{\small The \textquotedblleft 4-point gyrokinetic averaging\textquotedblright\ scheme employed in the charge deposition and push steps.  Ion centers are in green, their gyrokinetic charge rings in pink, and the grid points (in one of two planes) on which charge is deposited in blue.  Observe the overlapping grid accesses.
\label{fig:fourpoint}} 
\end{center}
\vspace{-15pt}
\end{figure}

{\bf Charge} performs a particle-to-grid interpolation in which particles deposit charge onto the charge grid using the 4-point gyro-averaging method (see Figure~\ref{fig:fourpoint}). 
Particles, represented by a 4-point approximation for a charged ring, may deposit charge onto as many as 32 unique grid memory locations and as few as eight.

In a shared memory environment, Figure~\ref{fig:fourpoint} visualizes the fact that threading over either particles or points on the charge ring can result in data hazards, which must be guarded by synchronization mechanisms in order to guarantee correct results.
Unfortunately, this challenge extends to SIMD architectures in which attempting to vectorize (or unroll) beyond 8-way (there are 4-points in the two enclosing poloidal planes) can produce intra-thread data dependencies.
Construction of thread (or lane) private copies of the charge grid can remedy the synchronization challenges, but can exacerbate cache and memory challenges. In a distributed memory environment, the charge deposition phase will also involve bandwidth-intensive MPI communication (a reduction).
For example, particles deposit charge on local grids consisting of non-overlapping zones and ghost zones. The charge on ghost zones needs to be merged to neighboring non-overlapping zones. 
In addition, if the domain is decomposed in toroidal dimension, the charge from neighboring toroidal sections needs to be merged. Additionally, the charge phase includes
a {\tt MPI\_Allreduce} to obtain a flux-surface-averaged charge for solving the so-called \textquotedblleft zonal flow\textquotedblright.

The {\bf Poisson, Field, Smooth} kernels solve the gyrokinetic Poisson equation, compute an electric field, and smooth the charge and potential with a filter on the grid, respectively. These three steps constitute purely grid-related work in which the floating-point operations and memory references scale with the number of poloidal grid points. As in the charge deposition phase, in a distributed memory environment, all three subroutines may involve MPI communication to 
update the values in the ghost zones. 

{\bf Push} interpolates the electric field onto particles, and using that field, advances particle phase space positions. In the push phase, the electric field 
values at the location of the particles are ``gathered'' and used for time-advancing their phase space coordinates. This step also 
requires reading irregular grid locations in memory (the electric field values) corresponding to the four bounding boxes of the 
four points on the ring, involving data reads from up to 32 unique memory locations. Fortunately, the operations in the push step are usually independent (hazard free), and are thus relatively easier to parallelize and vectorize.

In distributed memory environment, {\bf Shift} identifies particles being moved, copies them to a separated buffer and moves them to the appropriate processes.

In general, particle-related work such as {\bf charge}, {\bf push}, and {\bf shift} scales linearly with the number of particles. Grid-related work such as {\bf poisson, smooth} and {\bf field} scales with the number of 
poloidal grid points. Since particle to mesh ratio is in the order of 100 to 100,000 in PIC methods, the time spent in grid-related operations is a small component of the overall run time.

\subsection{Parallel Decomposition}\label{subsec:parallel}
GTC-P employs two levels of decompositions. First, a two-dimensional domain decomposition occurs in the toroidal dimension \emph{zeta} and the radial dimension \emph{psi} (see Figure 1).
In order to further increase MPI parallelism, a second level of decomposition over the particles is introduced. 
Within each subdomain, the particles are divided between several processes wherein each process owns a fraction of the total particles in that subdomain, as well as a private copy of the local grid to simplify the charge deposition.
The 2D domain decomposition and particle decomposition are implemented with MPI using three different communicators: a toroidal communicator, a radial communicator, and a particle communicator. 
The particles move between domains with nearest-neighbor communication (implemented with {\tt MPI\_Sendrecv}) within the toroidal communicator and radial communicator. To merge these private copies together, GTC-P performs a {\tt MPI\_Allreduce} within
the particle communicator. In gyrokinetic simulations, particles move much faster in the toroidal dimension than in the radial dimension at every time step. The messages 
transferred within the toroidal communicator are of much larger size than the messages exchanged within the radial communicator. To optimize performance, if possible, MPI ranks should be placed
such that the physical processors assigned to adjacent toroidal domains in the same toroidal communicator stay close together.

The domain decomposition in the toroidal dimension is trivial as the toroidal grid is uniformly discretized. However, with unstructured grid in circular geometry, the domain decomposition in the radial dimension is far more challenging.
We begin the process by partitioning a poloidal plane into non-overlapping domains with equal area. 
Assuming particle density is uniform, this partitioning divides all particles in one toroidal section equally across multiple processes. Next, the non-overlapping
  domain is extended to line up with the mesh boundary in the radial direction. Finally, the valid grid is extended on each side with ghost cells accounting for charge deposition with 4-point approximation.
In general, 3 to 8 ghost cells are sufficient. Though this discretization will result in more grid points for the processes close to the edge, the effect from load imbalance is small considering that 
the number of particles is much larger than the number of grid points (e.g., particles per cell number is much greater than 1).
In the case where more than 8 ghost cells are required for charge decomposition with the 4-point averaging, we can either send the off-region points explicitly to the neighboring
processes \cite{Wang12} or completely ignore those points. For the later case, benchmarking will be required to ensure that accuracy is preserved with the approximation.

The multiple levels of decomposition enables GTC-P to efficiently carry out \textquotedblleft true weak scaling study\textquotedblright\ where both the grid size and the number of particles are increased in proportion to the number of processors.
With a 2D domain decomposition and particle decomposition, GTC-P pushes the scalability of the PIC method to an extreme and allows scaling on the world's largest computing systems~\cite{Wang13}. Note that 
although the simulations are in 3D, a 2D domain decomposition is sufficient for weak scaling since the number of grid points in the toroidal dimensional remains constant for all plasma sizes due to Landau damping physics.

%======================================================================================
\section{Experimental Platforms}
%\MPar{LO: Reordered platforms to match tables/graphs}
In this paper, we examine the performance and optimization of GTC-P on six different systems.  We detail the machines here.

{\bf Edison:~} is a 5600 node Cray XC30 MPP at NERSC~\cite{EdisonWeb}.  Each node includes two 12-core Xeon (Ivy Bridge) processors running at 2.4GHz.  Each processor includes a 30MB L3 cache and four DDR3-1600 (the machine has since been upgraded to DDR3-1866) memory controllers which provide nearly 40GB/s of memory bandwidth.  Edison's nodes are connected together using Cray's Aries (Dragonfly) network which provides lower latency, higher bandwidth, and better scalability than torus or fat tree topologies.  As each node has two NUMA nodes (one per processor), we run one 12-thread process per NUMA node.  

{\bf Titan:~} is a GPU-accelerated Cray XK7 MPP at the Oak Ridge National Lab and is composed of 18,688 nodes connected via Cray's Gemini 3D torus like Blue Waters~\cite{TITANWEB}.  Unlike the similar Blue Waters, there are only 2 Interlagos CPU processors per node with the others being replaced by a PCIe-attached NVIDIA Kepler K20x GPU.  Each GPU, running at 733MHz, has 14 streaming multiprocessors each of 192 CUDA cores, 64 DP data paths, a 64KB SRAM which can be partitioned into a cache and a scratchpad, and 32K 32-bit registers.  Each GPU includes 6GB of GDDR5 memory.    The GPU presents programmers with dramatically different performance characteristics than CPU architectures as it favors computations that maximize parallelism, present streaming access patterns with easily expressible locality, and minimize off-node (off-GPU) communication.  

When using Titan, there are several programming styles which must mix MPI, OpenMP, and CUDA.  First, one can run one 16-thread, GPU-accelerated process per node.  This minimizes off-node computation and allows one to fully exploit the GPU.  However, this  demands the programmer optimize for NUMA in OpenMP --- a challenging prospect.  Second, one can run one 8-thread, GPU-accelerated process per node.  This has all the benefits of the aforementioned style, but side steps NUMA by constraining the process to a single NUMA node.  Unfortunately, this cuts CPU bandwidth and memory in half.  Finally, one may ignore the GPU altogether and run two 8-thread processes per node.

{\bf Mira:~} is an IBM Blue~Gene/Q supercomputer at the Argonne National Lab~\cite{MIRAWEB}. Mira is composed of 49,152 compute nodes connected via a high-performance custom network in a 5D torus.  Each node includes a 16-core (+1 system core) 1.6GHz SOC processor connected to 16GB of DDR3-1333 DRAM via two 128-bit memory controllers.  Each core is in-order, dual-issue, 4-way multithreaded, has a 4x64b SIMD FMA functional unit, and includes a private 16KB cache.  One must run at least two threads per core in order to dual-issue instructions.  All cores are connected via a crossbar to a 32MB L2 cache.  

{\bf Blue Waters:~} is a Cray XE6 MPP at the the National Center for Supercomputing Applications~\cite{BLUEWATERSWEB}.  It contains 22,500 Interlagos-based nodes connected by Cray's Gemini network into a 3D torus.  Each node includes four 8-core processors running at 2.3GHz in which each core has a private 16KB L1 data cache, each pair of cores shares a 256b-wide SIMD unit, each pair of cores shares a 2MB L2 cache, and all cores share a 32MB L3 cache.  Each processor is connected to 16GB of DDR3-1333 DRAM via a 128-bit memory controller.  As such, each compute node contains four NUMA nodes and programmers are motivated to run 4 processes of 8 threads per node.   Computation must be partitioned across NUMA nodes in order to maximize performance.  Each processor (NUMA node) thus provides about half the bandwidth and one-third the compute capability of a Blue~Gene/Q node, but has a much more cache.

{\bf Piz Daint:~} is a 5,272-node GPU-accelerated Cray XC30 MPP at the Swiss National Supercomputing Center (CSCS)~\cite{PIZDAINTWEB}.  Piz Daint is very similar to Titan with the caveat that rather than using Cray's Gemini 3D torus, nodes are connected via Cray's Aries dragonfly network.  Aries provides much higher bandwidth, fewer hops, and overall, better scalability than Gemini.  The other significant difference is the switch from Interlagos to Xeon (Sandy Bridge).  As a result, there is only one NUMA  node per compute node.  As such, one may simply run one 8-thread process per node and ignore any NUMA optimizations. Performance per NUMA node is broadly comparable to Edison.

{\bf Stampede:~} is a 6400-node Xeon Phi (MIC)-accelerated Infiniband cluster at TACC~\cite{STAMPEDEWEB}.  Each node includes two 8-core Xeon (Sandy Bridge) processors running at 2.7GHz and a PCIe-connected 60-core Xeon Phi co-processor.  Each Xeon Phi core is a dual-issue, in-order, 4-way multithreaded core capable of executing one 8-way SIMD FMA per cycle.  Each core has a private 32KB L1 cache and a coherent 512KB L2 cache.  Each Xeon Phi is attached to 8GB of GDDR5 memory providing 160~GB/s of streaming bandwidth.  In practice, the Xeon Phi and the Kepler GPU have comparable DRAM bandwidth, but the Xeon Phi has a more conventional cache hierarchy and can run existing MPI+OpenMP applications in native mode.  However, many of the programmer-friendly micro-architectural features of the commodity Xeon processors are not present in the Xeon Phi.  As a result, attaining high performance can be difficult.  In addition to native mode (host-only) performance experiments, we also examine the performance of Xeon/Xeon Phi clusters in symmetric mode.  In symmetric mode, hosts and accelerators are viewed as piers each running an MPI process.  Although simpler than offload mode and more capable of fully utilizing a heterogeneous node than native mode, load balancing between Xeon processes and Xeon Phi processes becomes a challenge.

%{\bf Babbage:~} is a 45-node Xeon Phi (MIC)-accelerated Infiniband cluster at NERSC\fix{reference}.  Each node includes two 8-core Xeon processors running at 2.6GHz and two PCIe-connected 60-core Xeon Phi co-processors.  Each Xeon Phi core is a dual-issue, in-order, 4-way multithreaded core capable of executing one 8-way SIMD FMA per cycle.  Each core has a private 32KB L1 cache and a coherent 512KB L2 cache.  Each Xeon Phi is attached to 8GB of GDDR5 memory providing 160~GB/s of streaming bandwidth.  In practice, the Xeon Phi and the Kepler GPU have comparable DRAM bandwidth, but the Xeon Phi has a more conventional cache hierarchy and can run existing MPI+OpenMP applications in native mode.  However, many of the programmer-friendly micro-architectural features of the commodity Xeon processors are not present in the Xeon Phi.  As a result, attaining high performance can be difficult.

{ \setlength{\tabcolsep}{1pt} \begin{table*}[!tb] \centering \footnotesize \begin{tabular}{c|c|cc|c|c|cc|cc}
                                      &	 {\bf Edison}           &	 \multicolumn{2}{c|}{\bf Titan}           &		 {\bf Blue}               &	 {\bf Mira}         &	 \multicolumn{2}{c|}{\bf Piz Daint}                   &	 \multicolumn{2}{c}{\bf Stampede}        \\		
                                   &	 {\bf XC30}             &	 \multicolumn{2}{c|}{\bf XK7}             &		 {\bf Waters}           &	 {\bf BGQ}         &	 \multicolumn{2}{c|}{\bf XC30}                          &	 \multicolumn{2}{c}{\bf MIC Cluster}        \\		
\hline%---------------------------------------------------------------------------------------------------------------------------------------------									
                                   &	 Intel                      &	 AMD                            &	 NVIDIA            &	 AMD                       &	 IBM                    &	 Intel                    &	 NVIDIA                                     &	 Intel                   &	 Intel                \\
Processor                    &	 Xeon                     &	 Opteron                   &	 Kepler                  &	 Opteron                  &	 BGQ                   &	 Xeon                   &	 Kepler                                     &	 Xeon                   &	 Xeon Phi        \\
\hline%---------------------------------------------------------------------------------------------------------------------------------------------									
Freq. (GHz)                   &	2.4   &	2.2   &	0.733   &	2.3   &	1.6   &	2.6   &	0.733   &	2.7   &	 1.1                        \\
D\$ per core (KB)         &	 32+256              &	 16+2048$^\dagger$           &	64   &	 16+2048$^\dagger$           &	32   &	 32+256              &	64   &	 32+256           &	 32+512                \\
Cores per CPU              &	12   &	8   &	14   &	8   &	16   &	8   &	14   &	8   &	 60                        \\
D\$ per CPU (MB)         &	30   &	8   &	1.5   &	8   &	32   &	16   &	1.5   &	20   &	 ---                        \\
\hline%---------------------------------------------------------------------------------------------------------------------------------------------									
CPUs/Node                   &	2   &	2   &	1   &	4   &	1   &	1   &	1   &	2   &	 1\\
DP GFlop/s                   &	460.8   &	140.8   &	1314   &	294.4   &	204.8   &	166.4   &	1314   &	345.6   &	 1056                \\
STREAM (GB/s)              &	 78$^*$               &	 31$^*$                        &	171   &	 62$^*$                          &	26   &	 38               &	171   &	78$^*$        &	 160                \\
STREAM/NUMA             &	39   &	15.5   &	171   &	15.5   &	26   &	 38              &	171   &	39             &	 160                \\
Memory (GB)                  &	 64$^*$              &	 32$^*$                       &	6   &	 64$^*$                           &	16   &	32   &	6   &	 32$^*$               &	 8                        \\
\hline%---------------------------------------------------------------------------------------------------------------------------------------------									
Network                        &	  Aries                 &	  \multicolumn{2}{c|}{Gemini}                                  &	 Gemini                           &	 5D        &	  \multicolumn{2}{c|}{Aries}                 &	  \multicolumn{2}{c}{Infiniband}        \\        			
and Topology                &	 Dragonfly          &	  \multicolumn{2}{c|}{3D torus}                             &	 3D torus                         &	 Torus     &	  \multicolumn{2}{c|}{Dragonfly}          &	  \multicolumn{2}{c}{Fat Tree}        \\        			
\hline%---------------------------------------------------------------------------------------------------------------------------------------------									
Compiler                       &	 Cray                  &	 Cray                           &	 NVCC                       &	 Cray                           &		XLC                   &	Cray                                   &	NVCC                  &	ICC     & ICC           \\
\end{tabular}
\caption{\small Systems evaluated in this paper.  $^\dagger$Each pair of cores shares a 2MB L2 cache.  $^*$NUMA. }
\label{tab:machines}
\end{table*}
}

\section{GTC-P on Modern Node Architectures}\label{subsec:opt}
%\MPar{LO: I moved the machine section earlier and split up GTC optimization - hopefully this makes sense?}The HPC community is moving to the era of multi- and many-core architectures with greatly increased parallelism on shared memory processors and as well as wider vectorization.
In order to harness the computing power of the emerging architectures, applications need to be carefully designed such that the hierarchy of parallelism provided by the hardware is fully utilized. 
In GTC-P, aside from inter-node parallelization implemented with MPI, we explore on-node scaling with shared memory thread-level parallelism implemented with OpenMP and in-core vectorization with SIMD intrinsics. The structure of array (SoA) layout is 
chosen for particle data to maximize spatial locality on streaming accesses. The data allocation is aligned to memory boundary to facilitate SIMD intrinsics.

For heterogeneous systems with GPU accelerators, we judiciously select computationally expensive kernels such as {\bf charge} and {\bf push} as well as the mandated {\bf shift} kernel to run on GPUs while leave the communication intensive 
subroutines such as {\bf smooth}, {\bf poisson} and {\bf field} on CPUs. GPU architectures exhibit multiple distinguishing computational constraints including limited-capacity high-bandwidth memory, limited coherence
support, low-overhead scheduling, and high-throughput vector computing. For instance, their small memory to core ratio restricts the use of memory replication as a technique to avoid data hazards in {\bf charge}. We also found it more efficient to redundantly recompute values rather than precompute and load from memory.  
Additionally, loop/computation fusion was implemented to further reduce memory usage for auxiliary arrays in {\bf push}. The GPU's relatively small cache capacity makes it difficult to exploit locality to avoid expensive memory accesses. We addressed this limitation previously using cooperative computation to capture locality for
co-scheduled threads (thread blocks)~\cite{MII11}. In this paper, we also explore different techniques for explicit management of the GPU shared memory. Like CPUs, GPUs benefit greatly from sorting particles based on their cell
association. Sorting helps in reducing the  temporal reuse distance and aids the cache hierarchy in capturing locality. We exploit the GPUs low-overhead hardware scheduler to load-balance computation using fine-grained tasks (one particle computation per thread). Ultimately, the high overhead of moving data between the GPU and CPU host motivated us to keep particle persistently allocated on
the GPU memory. Only shifted particles are moved to/from the host during communication phases. 

In this section, we detail our efforts in optimizing GTC-P kernels on multi-core CPU and many-core GPU.

%\subsection{Charge}
{\bf Charge:}~
Since modern computer architectures rely on data locality and regular computation to tolerate access latency, it is extremely challenging to optimize the {\bf charge} kernel 
due to the random memory access, potential fine-grained data hazard and low computational intensity characteristics. In order to improve data locality for grid access in {\bf charge} and {\bf push}, 
the particles are binned in the radial dimension. The binning frequency is an adjustable parameter given by the users. 

In multithreading environment, two particles operating by two different threads may access 
the same grid point and cause a read-after-write data hazard. Thus, the data access must be either guarded with a synchronization mechanism or redirected to a private copy depending on the availability of memory resources.
On CPUs with up to hundreds of threads, the grid replication strategy on a per thread basis is usually used to address the issue of fine-grained data hazards \cite{Madduri11,Wang13}. The charge densities on each private copy are summed at the end of the charge phase, where the order of summation is carefully chosen to avoid synchronization. 
Note for large size plasma simulation, the grid replication strategy is only applicable with 2D domain decomposition. 

On GPUs with thousands of threads, the grid replication strategy is no longer applicable even with particle binning in all dimensions. This is because in gyrokinetic PIC simulation, each particle is actually a gyroparticle represented by 4 points on a ring of 
gyroradius. Even with perfectly sorting, the influential grid points from a set of particles of a single cell will spread up to 16 radii.
In a circular geometry, each thread may require as much as 2 MB private memory to store the local grid information such as charge density and field --- cost prohibitive with thousands of threads.
This motivates us to use an update binning algorithm to implement charge deposition on the GPU. Besides binning gyroparticles according to their coordinates periodically, we also bin the points of all gyroparticles at every time step. 
This additional points binning step simplifies the gyrokinetic PIC algorithm to the standard PIC algorithm. As a result, a local grid private to each thread spans only up to 1 radii instead of 16.
Due to reduced ghost zones, the additional point binning significantly reduces the temporary memory storage for hazard-free charge deposition. Hence, we can use the fast shared memory for charge deposition.

Specifically, the binned points are carefully organized such that all points belonging to the same super-cell are processed by the same CUDA thread iteratively and the points in two adjacent super-cells are processed by 
two consecutive CUDA threads. Every thread block holds a local copy of the grid on shared memory for charge deposition relying solely on fast 
shared memory atomic operations. Furthermore, atomic operations could be totally avoided by providing 
two copies of the partial grid in shared memory, one for charge deposition by threads with even IDs and another with odd IDs. 
The charge in shared memory is added to global memory at once at the end. Given the extensive use of shared memory, we configure the GPU to favor shared memory.

One drawback of the above algorithm is when the number of points in different super-cells has large variations, leading to potential issues of memory waste and load 
imbalance. As we discussed earlier, in a Monte Carlo type particle simulation, the particles are loaded with importance sampling such that the particle number density is
proportional to the distribution function $f$. In the $\delta f$ method, this translates to the particle number density is proportional to the background 
distribution function $f_0$. As we consider uniform background density currently, we should expect relatively uniform particle number
density in each cell. However, since GTC-P uses magnetic coordinates, the Jacobian from the coordinate transformation (from magnetic coordinates to Cartesian coordinates) needs be taken into account.
Consequently, the particle number density will be proportional to the Jacobian
%\begin{equation}
$ J = {(1 + r \cos(\theta))}^2,$
%\end{equation}
where $r$ and $\theta$ is the radial and poloidal coordinates. This issue can be addressed by developing a non-uniform grid on each poloidal ring such that the number of points on the new grid has small variations.
Note that the non-uniform grid discretization is only used for the purpose of binning points.

The updated binning algorithm works especially well on NVIDIA's previous Fermi architecture, where double-precision atomic increments are expensive.
The subsequent introduction of the NVIDIA Kepler architecture (employed in Titan and Piz Diant) enabled fast double-precision atomic increment (implemented via a compare-and-swap operation).
As a result the performance of the charge algorithm is limited by the number of active thread blocks in a streaming multiprocessor. While Kepler preserves the same amount of shared memory as the earlier version Fermi, we observe that
the updated binning algorithm is less appealing on Kepler. Alternatively, the previous implementation using cooperative computation \cite{Madduri11} that uses global atomic write performs  well if designed properly.
Specifically, computation is organized so that CUDA threads access successive particles when reading, but is reorganized so that threads work together on one charge deposition (of up to 32 neighboring grid points) when writing. 
The reorganization is through transposing the data write in shared memory. Given the limited use of shared memory in this case, the GPU is configured to favor the L1 cache. \newline

%\subsection{Poisson/Field/Smooth}
{\bf Poisson/Field/Smooth:}~
The grid-related kernels are run on the CPUs for both homogeneous and heterogeneous systems. This is mainly because simulations typically employ high particle number
densities.  As a result, the time spent in grid-based kernels is substantially lower than the particle-related kernels. In addition, due to two-dimensional domain
decomposition, all grid-based kernels involve extensive MPI communication to update the values at ghost zones. 

The poloidal planes in tokamaks are in circular geometry, thus the grid-related kernels usually include irregularly nested loops in which {\emph psi} is the outer loop and 
{\emph theta} is the inner loop. On large tokamak devices, the outermost domain has little variability in the outer loop, but large variability in the inner loop. Simple
thread parallelization of the outer loop among threads is not sufficient for multi-core architectures with thread numbers varying up to 64. The main optimization we used for the grid-based kernels is manually flattening the nested loop into a single loop to expose more thread-level parallelism. 
Note the OpenMP {\tt collapse(2)} clause is not applicable here as it requires that the nested loops form a rectangular iteration space.
\newline

%\subsection{Push}
{\bf Push:}~
The {\bf push} kernel involves two operations: gather and update. The key optimization we employ is loop fusion to avoid the write and read of temporary particle data. The loop fusion optimization also minimizes parallel thread 
fork-join overhead and increases computational intensity of the kernel. The data locality for grid access is improved by binning the particles in the radial dimension. For large tokamak devices, binning in the radial dimension only may lead to large cache or even TLB misses in CPUs,
since the outermost rings usually consist of a large number of grid points. On GPUs, it is important to place the electrostatic field on CUDA's texture memory 
to maximize the read bandwidth from non-coalescing memory access. \newline

%\subsubsection{Shift}
{\bf Shift:}~
This kernel streams through the particle array with unit access, identifies the particles to move, packs them in a separated buffer and sends/receives them to/from the neighboring processors with message passing.
On CPUs, the performance of this kernel sensitively depends on two factors. One is the network performance of the system, in particular for nearest neighbor point-to-point communication. 
The second factor is the effect of caches misses in packing the particles in a separated buffer. On heterogeneous systems with GPUs, particle packing is implemented on the GPU while the message passing remains on the CPU. Besides the network performance of the system, the performance 
of {\bf shift} kernel will depend on how well the GPU handles the packing process and the performance of data transfer between the CPU and the GPU via the PCIe bus.

In the previous GPU implementation of particle packing~\cite{Madduri11},
each thread block maintains small shared buffers that are filled as it traverses its subset of the particle array. Particles are sorted into three buffers for left shift, right shift and keep buffer. Whenever the local buffer is exhausted,
 the thread block atomically reserve a space in a pre-allocated global buffer and copy data from the local buffer to the global one. Recall that the normal iterative shift 
algorithm on CPUs uses array of structure data layout to pack the data for message passing, the data is transposed while flush to the global buffer. Then the data is copied back
to the CPU where the message passing is executed. Upon completion, the host transfers a list of incoming particles to the GPU, where unpacking involves filing clustered holes in the particle arrays and transposing the data back to the GPU data structure 
of arrays layout. With data transposes and atomic operations, maximizing parallelism and attaining good memory coalescing make porting shift kernel extremely challenging on GPUs. 
This strategy works well on the NVIDIA Fermi Architecture, even though it slightly limited the number of active thread blocks (thread block occupancy). 

With the introduction of  the NVIDIA Kepler architecture with Streaming Multiprocessor Architecture (SMX), the core count per multiprocessor increased and the core speed decreased, while the amount of shared memory is preserved similar to earlier generation Fermi. As such, we redesigned areas of the code that stress the use of the shared memory and limit occupancy including the shift routine.  
 In the latest version of the code, the shift and sort functionality are  simultaneously executed in the same phase. Relying on the fast sorting algorithm provided by Thrust library \cite{Thrust}, particles are sorted into three
buffers for the left shift, right shift and keep buffer in the GPU's global memory. By modifying the normal iterative shift algorithm on the CPUs to pass message with structure of array data layout instead of array of structure, the data transpose operation is not required. Compared with the previous approach, both reliance on the GPU's shared memory and data transpose operations are reduced. \newline

%\subsection{Intel Xeon Phi Implementation}  \label{subsec:mic}
{\bf MIC-Specific Implementation:}~
The Intel Xeon Phi supports several execution models, including native, offload and symmetric mode. As its programming environment is based on C/Fortran languages and MPI/OpenMP standards,
it is reasonably straightforward to port the code onto the MIC processor. However, attaining high performance can be difficult on some applications. This is mainly due to several differences in the design of the processor 
cores on Xeon Phi \cite{Rahman13}. First, MIC supports much larger number of threads than the Xeon processor while the performance of each thread is much lower than that of the Xeon processor. The MIC also has wider
vector units. Thus, high performance can only be achieved if the applications utilize all cores and vector units effectively. Additionally, although the peak performance of Xeon Phi is much higher than the Xeon processor, the memory bandwidth per core is not. As such,
the lower flop per byte on MIC may require choosing a different data structure and algorithms for solving the same problem. In this paper, we share our early experience in porting and optimizing GTC-P code onto Intel Xeon Phi in native mode and symmetric mode. As PIC algorithm involves \textquotedblleft gather\textquotedblright\ and \textquotedblleft scatter\textquotedblright\ operations that features
random memory access and potential data hazards, our studies show some challenges and opportunities of using Intel Xeon Phi for irregular applications.

The initial performance of the code on Xeon Phi in native mode was 2x-4x slower compared with the same code running on host CPU.  
%\MPar{(why?  where was time going??? Kamesh has pulled out our email conversation about this back to 09/03/2013. Here is a short summary of our observation.)}
As the number of threads per core increased from 1 to 4, {\bf shift} performance did not scale as the number of OpenMP threads were increased, instead the runtime grew linearly with the number of threads. Further investigation concluded that this was due to the \textquotedblleft malloc\textquotedblright~and \textquotedblleft free\textquotedblright~operations performed by each thread.
With 240 threads, the calls for these two operations were serializing and thus causing poor scalability. The first optimization we applied is to replace heap memory created by each thread within the OpenMP parallel region with pre-allocated memory buffer on {\bf shift}.
%For he {\bf push} kernel involves two operations: gather and update.
%\MPar{LO removed this sentence: Also note that the performance degradation due to the serialized \textquotedblleft malloc\textquotedblright~and \textquotedblleft free\textquotedblright~has been improved significantly in the current
%version of the Intel compiler (Intel C/C++ compiler 13.0).}
 On multi-core CPUs and many-core GPUs, we use loop fusion to increase computational intensity and to minimize parallel thread fork-join overhead for {\bf push}.
On Intel Xeon Phi, we find that loop fission boosts the performance by 20\%. Specifically, we implement the gather and update operations with two separate loops. For the gather loop, in order to decrease the memory access latency
due to \textquotedblleft gather\textquotedblright\ operation, the neighboring grid information is prefetched to level 1 cache. In addition, the grid memory read is implemented with vector intrinsics. 
The update loop does not include any irregular operation and can potentially have high performance on Intel Xeon Phi. We thus use {\tt \#pragma omp for simd} to enable SIMD instructions. 
Considering that the synchronization between the large number of threads in Xeon Phi could be expensive, we use the grid replication strategy, the same as we choose for multi-core CPU, to address the issue of fine grained data hazard in {\bf charge}.

In symmetric mode execution, MPI processes can run fully on both host and co-processor. However, it has been demonstrated that the latency and the bandwidth can vary in an order of magnitude depending on the pair of communication devices involved \cite{Karpusenko14}. 
The multiple levels of decompositions in GTC-P provides great flexibility in mapping the MPI ranks to physical devices with relatively optimal communication bandwidth. 
To enable message passing through only high bandwidth channels, we turn on particle decomposition and confine all MPI ranks among the same node to a particle communicator. 
With particle decomposition among the MPI ranks of the same node including the host CPUs and the MICs, we can further enhance the performance by running the communication intensive grid-based subroutines only
on the host CPUs. The grid-based charge density and electrostatic field can be copied from and to the co-processors with {\tt MPI\_Reduce}  and {\tt MPI\_Scatter} within the particle communicator. 
Since the collective communication only involves message passing inside a node, this avoids communications that potentially involve varying latency and bandwidth. 
Depending on the performance of the particle-based kernels (e.g., {\bf charge}, {\bf push}) on different devices, we distribute the particles between them such that the work on different devices is well balanced. 

We also highlight that the current GTC-P symmetric mode implementation is similar to the GPU implementation and future offload mode implementation in that only particle-based kernels are running on accelerators.
Unlike the offload approach in which the host CPUs and the accelerators are running in a staggered pattern, both the host CPUs and the MICs are running concurrently in symmetric mode.

%======================================================================================
\section{Results and Analysis} \label{sec:results}
In this section, we demonstrate the performance of GTC-P on the six evaluated supercomputing systems using weak scaling on different plasma sizes.  
In contrast to the weak scaling studies by our previous studies~\cite{Madduri11,Ibrahim13} and by other particle-based fusion codes \cite{Meng13} where the grid size was kept constant but the number of particles was scaled, our 
experiments represent a true weak scaling study of both the grid size and the number of particles. This is important to effectively address the outstanding open issues in fusion plasma physics such as the scaling of the energy confinement time with system size.
A summary of our recent ITG simulations using the highly efficient code is presented in \S \ref{sec:itg}.

\subsection{Configuration}
A GTC-P simulation is described by several important numerical parameters: the toroidal grid size \emph{ntoroidal}, the poloidal grid size \emph{mgrid}, and the average particle 
density as measured in the ratio of particles to grid points \emph{micell}; where the total number of particles in a simulation is $ntoroidal \times mgrid \times micell$. 
In order to demonstrate the viability of our optimizations across a wide variety of potential simulations, we explore four different grid problem sizes, 
labeled \textit{A, B, C, D}. Grid size $A$ and $B$ correspond to the majority of existing tokamaks in the world, $C$ 
corresponds to the JET tokamak, the largest device currently in operation \cite{JET}, and $D$ to ITER, the largest device currently under construction in France \cite{ITER}.
Table~\ref{tab:gridsizes} lists the poloidal grid sizes and the corresponding grid and particle related memory requirement in one toroidal domain for different plasma sizes.
Observe that when moving to a plasma device of one size larger (with constant \emph{micell}), e.g., $A$ to $B$, the poloidal grid size and number of particles in one toroidal domain increase by 4$\times$.
As discussed earlier, a production simulation generally consists of only fixed number of poloidal planes (e.g., 64), wrapped around the torus for all plasma sizes due to Landau damping physics. As as result,
in our weak scaling study, the number of processors increases 4$\times$ (instead of 8$\times$ as in some 3D codes) as we move to a plasma one size larger. In all of our experiments below,
we choose particle number density $micell=100$. There are two reasons of choosing a single particle number density in our current performance study. First, $micell=100$ is a typical 
number in all production runs using $\delta f$ method. Second, as suggested in earlier studies \cite{Madduri11,Wang13}, high particle density leads to improved parallel efficiency. Thus,
the results in this study will establish a performance base line for subsequent higher resolution simulation.

{
\setlength{\tabcolsep}{5pt}
\begin{table}[!tb]
\centering
\begin{tabular}{rrrrr}
\hline
{\bf Grid Size}	 &{\bf A}     & {\bf B}	& {\bf C}& {\bf D} \\\hline
\emph{mpsi}	& 90    & 180	& 360	& 720  \\
\emph{mthetamax}	&640 	  & 1280& 2560& 5120\\\hline
\emph{mgrid} (grid points per plane) & 32449 & 128893	& 513785 & 2051567\\
\emph{chargei} grid (MB)$^\dagger$ & 0.5 & 1.97 & 7.84 & 31.30			\\
\emph{evector} grid (MB)$^\dagger$ & 1.49 & 5.90 & 23.52 & 93.91		 \\
Total particles \emph{micell}=100 (GB) 	& 0.29 & 1.16	& 4.64	& 18.56		\\\hline
\end{tabular}
\caption{\small The GTC-P numerical settings for different plasma sizes. The grid and particle memory requirements are for one toroidal domain only. A simulation
typically consists of fixed number of toroidal domains, e.g., 64. \textquotedblleft mpsi\textquotedblright\ is the number of grid points in the radial dimension and \textquotedblleft mthetamax\textquotedblright\ is the number
of grid points in the poloidal dimension at the outermost ring. }
\label{tab:gridsizes}
\end{table}
}

\subsection{Performance Evaluation and Analysis}
Since the proper placement of MPI ranks to physical nodes/processors is important to the network performance, we have applied optimized 
placements on each system if applicable. In gyrokinetic simulations, because the parallel velocity is much larger than the perpendicular velocity (parallel and perpendicular to the magnetic field), 
at each time step, the number of particles moving out of the local domain in the toroidal dimension is 10$\times$ more than in the radial dimension. We thus place the MPI ranks to favor the toroidal communicator. 
Generally, this can be achieved by assigning consecutive MPI rank numbers for processes within each toroidal communicator.
On a Blue Gene/Q system with 5D torus network, we further group two or three torus dimensions to match 64 for an optimized placement layout by
setting the environmental variable RUNJOB\_MAPPING. The network resources are shared among all users on Titan and Blue Waters.
leading to large performance variation for the same experiment. In order to minimize this effect, we run the same experiment multiple times and record the best result. 
On Blue Waters, we further select a set of convex shaped computing nodes through job scheduler for optimal network performance.
The similar convex node selection is not available on Titan by the time the data was collected. 

\subsubsection{One Process per NUMA Domain}
NUMA nodes are increasingly viewed as the standard unit of computation.  When subtracting for age bias, most NUMA nodes has a comparable amount of memory and bandwidth.
As such, in our first series of experiments, we choose to run with one MPI process per NUMA node and thereby eliminate the need for NUMA optimizations. The 
configuration for each platform is shown at the top of Figure~\ref{fig:weakscale}, showing the MPI ranks $\times$ OpenMP threads, as well as MIC in native mode and GPU experiments. 

To fit the largest problem size D on Piz Daint, a 5,272 compute nodes system, we choose 4096 MPI partitions with 64-way parallelization
in both toroidal and radial dimensions. Our weak scaling study starts from A size problem using total 64 MPI processes with
64-way partitioning in the toroidal dimension and 1-way partitioning in the radial dimension. While the number of partitioning in the toroidal dimension remains unchanged in the weak scaling study, we increase the radial partitioning by 4$\times$ as we move to each larger plasma size. For all experiments with 1 MPI per NUMA node, we do not apply particle decomposition. 

{
\setlength{\tabcolsep}{2pt}	
\begin{figure*}[!tb]
\centering
\footnotesize
\begin{tabular}{rccccccccc}
\multicolumn{10}{c}{\bf \textcolor{blue}{1 MPI Process per NUMA Node}} \\
\hline
& \multicolumn{6}{c}{\bf --- CPU-only Configurations ---} & \multicolumn{1}{c}{\bf MIC-only} & \multicolumn{2}{c}{\bf CPU+GPU}\\
& {\bf Edison} & {\bf Titan} & {\bf Blue W.} & {\bf Mira} & {\bf Piz D.} & {\bf Stamp} & {\bf Stamp} & {\bf Titan} & {\bf Piz D.} \\ \hline
MPI Procs & 2 & 2 & 4 &  1 & 1 & 2 & 1 & 1 &  1 \\\hline
OMP Threads & 12 &8 & 8 &  64 & 8 & 8 & 240 & 8 &  8 \\\hline
Accelerators & ---& ---& ---&--- & ---& ---&  ---& +K20x & +K20x \\ \hline
\multicolumn{10}{c}{}\\
\end{tabular}

\begin{minipage}[t]{2.10in}\centering\par{{\bf A (MPI ranks: 64)}}\end{minipage}
\begin{minipage}[t]{2.10in}\centering\par{{\bf B (MPI ranks: 256)}}\end{minipage} 
\begin{minipage}[t]{0.80in}\centering\par{{\bf}}\end{minipage} \\
\begin{minipage}[t]{.49\columnwidth}\centering\includegraphics[width=\textwidth]{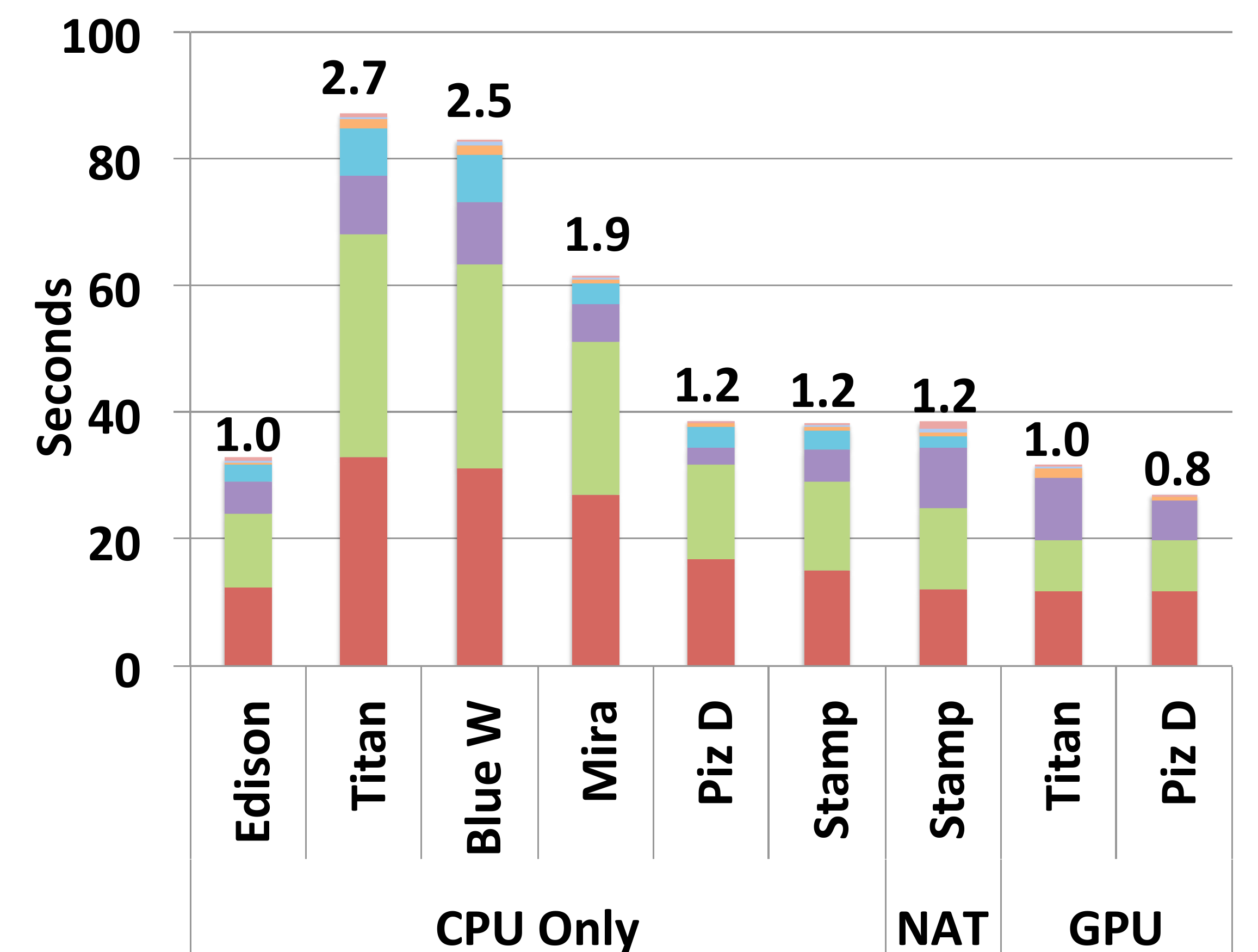}\end{minipage}
\begin{minipage}[t]{.49\columnwidth}\centering\includegraphics[width=\textwidth]{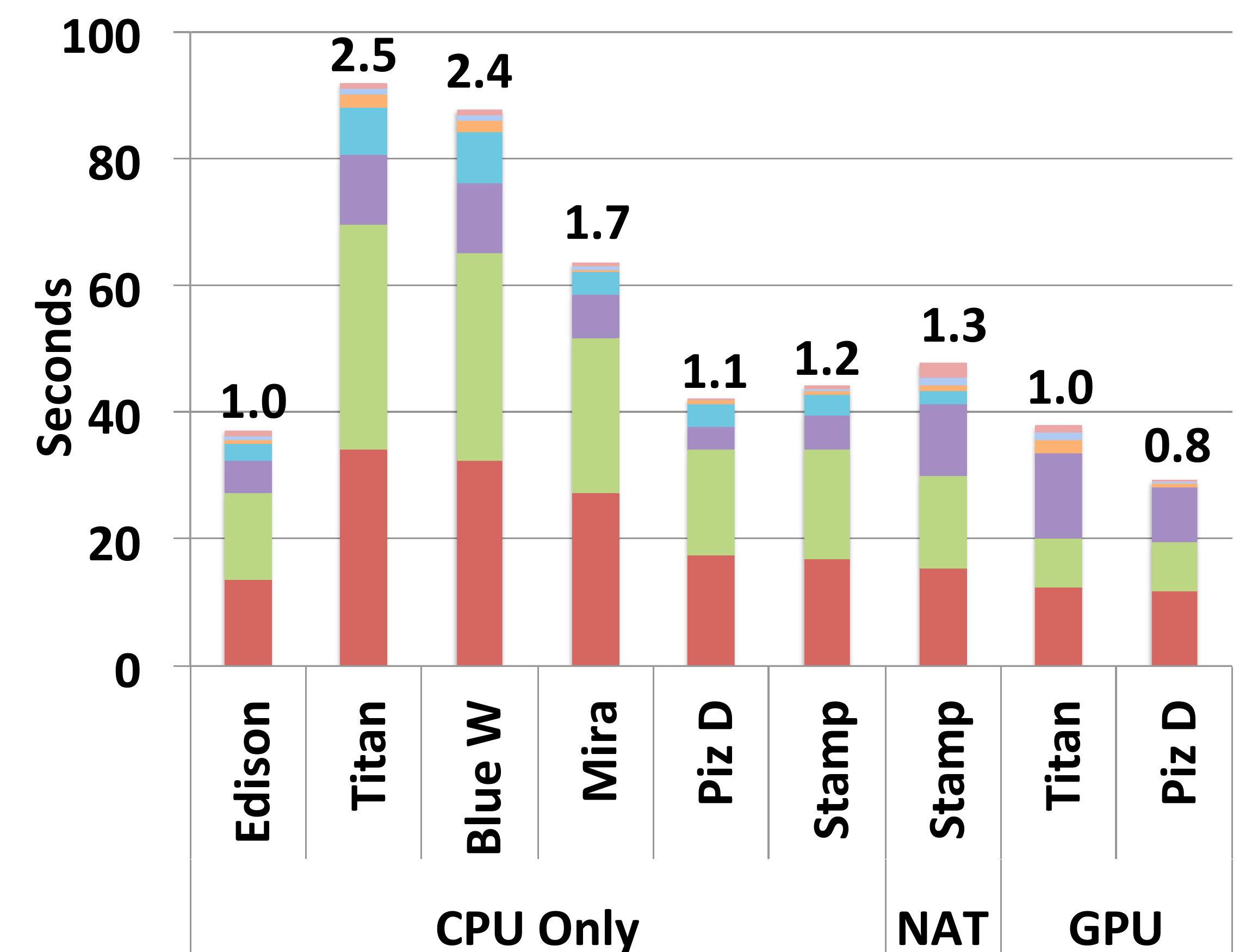}\end{minipage}
%\begin{minipage}[t]{0.80in}\centering\includegraphics[width=.99\textwidth]{../figure/legend2.pdf}\end{minipage}\\
\begin{minipage}[t]{0.40in}\centering\par{{\bf}}\end{minipage} \\
\begin{minipage}[t]{0.80in}\centering\par{{\bf}}\end{minipage} \\
\begin{minipage}[t]{2.10in}\centering\par{{\bf C (MPI ranks: 1024)}}\end{minipage}
\begin{minipage}[t]{2.10in}\centering\par{{\bf D (MPI ranks: 4096)}}\end{minipage}
\begin{minipage}[t]{0.80in}\centering\par{{\bf  }}\end{minipage} \\
\begin{minipage}[t]{.49\columnwidth}\centering\includegraphics[width=\textwidth]{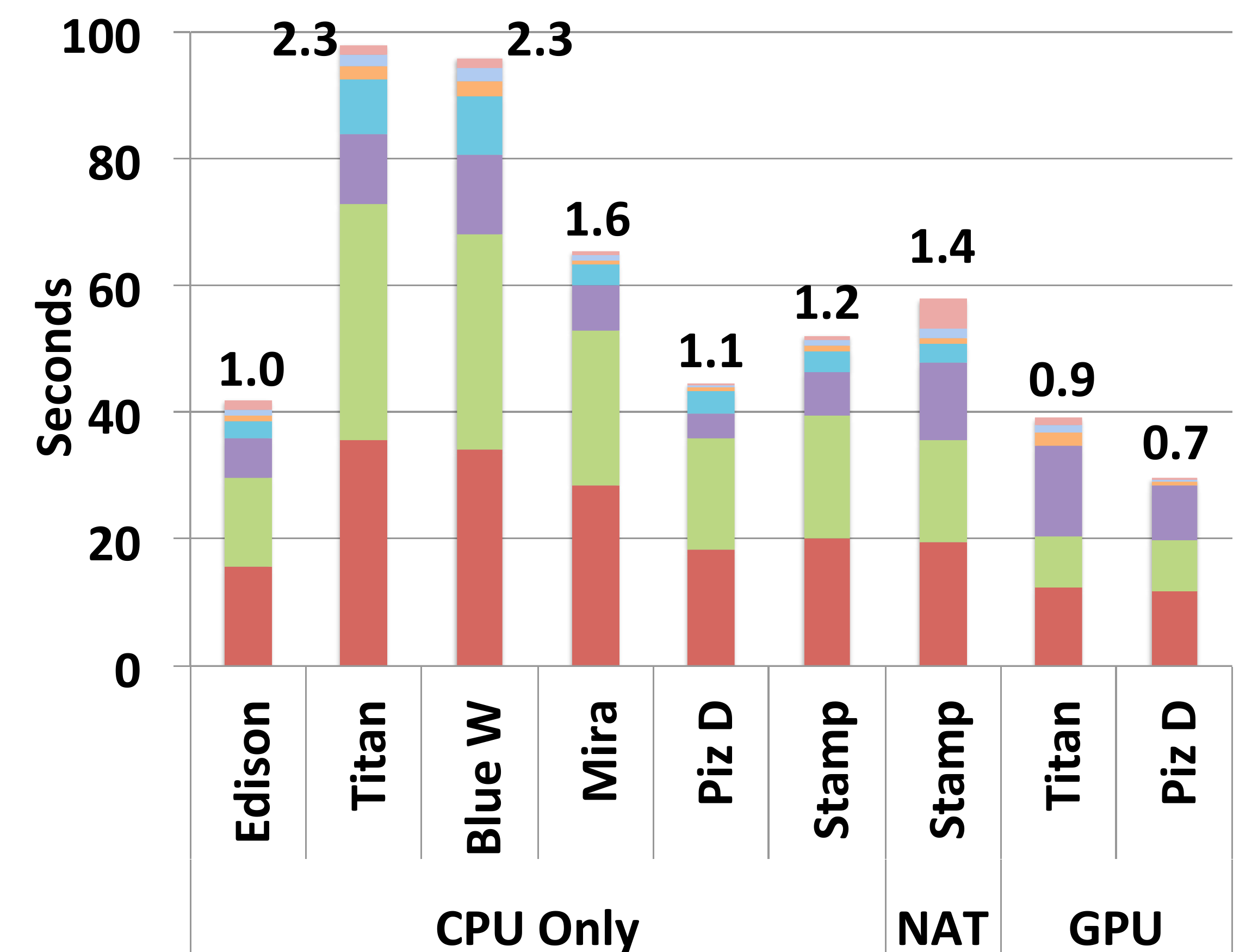}\end{minipage}
\begin{minipage}[t]{.49\columnwidth}\centering\includegraphics[width=\textwidth]{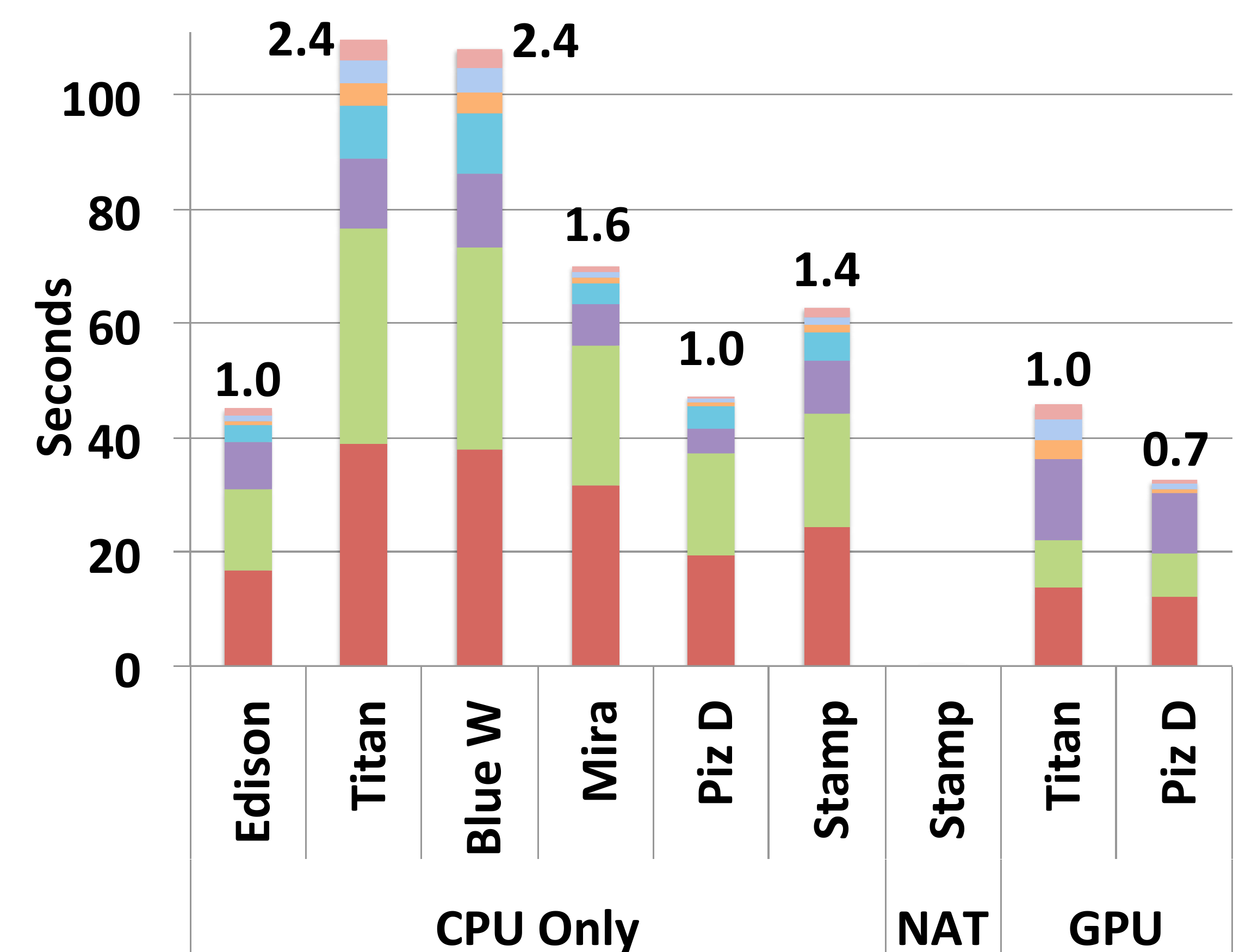}\end{minipage}
\begin{minipage}[t]{0.40in}\centering\par{{\bf}}\end{minipage} \\
\begin{minipage}[t]{3.50in}\centering\includegraphics[width=.99\textwidth]{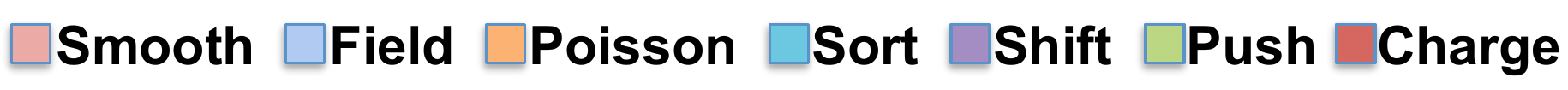}\end{minipage}

\caption{\label{fig:weakscale}
\small GTC-P Weak scaling run times for 100 time steps using plasma sizes A--D with 1 MPI process per NUMA  node. Table presents run configurations showing MPI ranks $\times$ OpenMP threads, as well as MIC in native mode and GPU experiments. In all experiments, the number of particles per process $\sim$3,235,900. Due to system issues, we were unable to conduct 4096 node native mode simulations on Stampede. Run time relative to Edison is shown above each bar (times greater than 1.0 designate a slowdown).}
%\end{center}
\end{figure*}
}

Figure \ref{fig:weakscale} presents the breakdown wall-clock time for 100 time steps on each kernel for the weak scaled plasma sizes A--D on our evaluated systems.
Note that on GPUs, the {\bf shift} and {\bf sort} routines have been merged to one phase in our new GPU-based shift kernel, thus their total wall-clock time are represented by {\bf shift} time only.

Observe, performance on the CPU-based architectures is generally correlated with the DRAM STREAM bandwidth per NUMA node as shown in Table~\ref{tab:machines}.
However, we observe that the overall run time on MIC-only and GPU-accelerated systems is substantially greater than the STREAM-predicted run time of 25\% of Edison.
There was no speedup on the MIC-only native mode performance over Edison on any function with {\bf shift} being substantially slower.
Conversely, on the GPU-accelerated systems, there was a substantial speedup on the easily parallelized and compute intensive push despite its data locality challenges with comparable performance on {\bf charge} and slowdowns on {\bf push}.  It is likely the data locality, synchronization, and PCIe challenges of these operations impede the performance of the GPU.

On the CPU systems, when weak scaling to larger size plasmas, the time spent in {\bf charge} increases for two reasons. First, there 
are load imbalance issues for charge interpolation on the 2D poloidal plane. Although the number of particles is well-balanced among different processes, the particle interpolation
close to the outer edge of the 2D will suffer larger cache- or even TLB misses. The problem is exacerbated on larger sized devices where the number of grid points at the edge increases. Second, as we discussed in \S \ref{sec:GTCP}, {\bf charge} includes
two type of communications, a point-to-point communication that merges values at ghost cell, and a collective communication that obtains flux-surface-averaged charge for solving \textquotedblleft zonal flow\textquotedblright. 
The time spent on both communications increases as the plasma size grows.
By examining the code using performance tools such as CrayPat on several Cray systems (i.e., Titan, Blue Waters and Piz Daint), we confirm that more L1, L2 and TLB cache misses occur for processors assigned to the domain close to the edge for C and D size plasmas.
Although MPI communication time increases from A to D as expected, the time is mostly spent at the synchronization point before the collective communication, thus suggesting that load imbalance is accounting for the major overheads in {\bf charge} kernel. 

The load imbalance caused by cache misses also accounts for the performance behaviors of {\bf push} as its operation only involve random memory access. 
The performance of {\bf shift} sensitively depends on the network performance. Therefore, the time increase from A to D for {\bf shift} can be used as an indicator of the network performance of a system.  The time spent on the grid-based subroutine has increased with large size devices. This is due to the increasing load imbalance on large plasma sizes from the radial domain decomposition in circular geometry. We observe that the time increase on Titan and Blue Waters is especially significant, where the time is mostly spent on 
the point to point communications in the radial dimension to update the ghost cell values. This issue could be alleviated by placing the MPI ranks to physical nodes to favor the radial communicator. However, this will affect the performance of {\bf shift} in toroidal communicator. 
The {\bf push} run time remains relatively flat from A--D plasmas as expected.

{
\setlength{\tabcolsep}{2.0pt}
\begin{table*}[!tb]
\begin{minipage}[b]{1.0\textwidth}\centering
{\footnotesize
\begin{tabular}{c|ccc|cc|ccc}
\multicolumn{9}{c}{{\bf Titan (1 MPI$\times$8 OpenMP + K20x per node)}} \\ \hline
\multicolumn{1}{c|}{} & \multicolumn{3}{|c|}{{\bf Charge}} & \multicolumn{2}{c|}{{\bf Push}} & \multicolumn{3}{c}{{\bf Shift+Sort}} \\
Problem & PCIe     & GPU Kernel & CPU Comm. &    PCIe       & GPU Kernel   & PCIe          & GPU Kernel & CPU Comm. \\ \hline
A100 &  0.05      &  11.65       &  0.30             & 0.15         &  7.73           & 4.25          & 1.99      &  3.70 \\ \hline
B100 &  0.05      &  11.48       &  0.71             & 0.17         &  7.57           & 4.34          & 3.90      &  5.18 \\ \hline
C100 &  0.05      &  11.48       &  1.00             & 0.18         &  7.47           & 4.36          & 3.89      &  6.07 \\ \hline
D100 &  0.05      &  11.38       &  2.71             & 0.18         &  7.29           & 4.35          & 3.86      &  6.30 \\ \hline
\multicolumn{9}{c}{} \\
\multicolumn{9}{c}{} \\
\multicolumn{9}{c}{{\bf Piz Daint (1 MPI$\times$8 OpenMP + K20x per node)}} \\ \hline
\multicolumn{1}{c|}{} & \multicolumn{3}{|c|}{{\bf Charge}} & \multicolumn{2}{c|}{{\bf Push}} & \multicolumn{3}{c}{{\bf Shift+Sort}} \\ 
Problem & PCIe    & GPU Kernel   & CPU Comm.   & PCIe         & GPU Kernel    & PCIe    & GPU Kernel & CPU Comm.   \\ \hline
A100 &  0.04      &  11.65         & 0.15 & 0.09         &  7.72        & 2.94          & 1.95   &  1.52  \\ \hline
B100 &  0.04      &  11.48         & 0.29 & 0.10         &  7.56        & 3.00          & 3.79   &  1.71  \\ \hline
C100 &  0.04      &  11.48         & 0.41 & 0.10         &  7.47        & 3.00          & 3.77   &  1.83  \\ \hline
D100 &  0.04      &  11.40         & 0.55 & 0.10         &  7.29        & 3.01          & 3.76   &  3.68  \\ \hline
\end{tabular}
}
\end{minipage}
\caption{\label{tab:weakscale2}\small Breakdown time (sec) for {\bf charge}, {\bf push} and {\bf shift+sort} \
using CPU+GPU on Titan and Piz Daint in Figure \ref{fig:weakscale}. }
\end{table*}
}

The GPU design relies on massive number of threads to hide the memory access latency (as opposed to hierarchies of caches on commodity CPUs). 
Thus, compared with the CPU-only results on Titan and Piz Daint (Figure \ref{fig:weakscale}), our CPU/GPU hybrid implementation results in a significant 2$\times$ and 3$\times$ speedup on {\bf charge} and {\bf push} kernels respectively.
In order to understand the GPU performance in more details, we measure the time spent in PCIe, in GPU execution and in CPU communication, respectively.
Table \ref{tab:weakscale2} lists these breakdowns for three kernels, {\bf charge}, {\bf push} and {\bf shift} from Figure \ref{fig:weakscale}.

{
\setlength{\tabcolsep}{1pt}
\begin{figure*}[!tb]
\centering
\footnotesize
\begin{tabular}{rcccccccccc}
\multicolumn{11}{c}{\bf \textcolor{blue}{1 MPI Process per Compute Node}} \\
\hline
& \multicolumn{6}{c}{} 							& {\bf MIC-}	& {\bf CPU+}	& \multicolumn{2}{c}{\bf CPU+}\\
& \multicolumn{6}{c}{\bf --- CPU-only configurations ---}	&  {\bf only}	& {\bf MIC}	& \multicolumn{2}{c}{\bf GPU} \\
& \bf{Edison} & \bf{Titan} & \bf{Blue W.} & \bf{Mira} & \bf{Piz D.} & \bf{Stamp} & \bf{Stamp} & \bf{Stamp} & \bf{Titan} & \bf{Piz D.} \\ \hline
MPI Procs & 1 & 1 &  1 &  1 &  1 &  1 &  1 &  1+1 &  1 &  1\\\hline
OMP Threads & 24 & 16 &  32 &  64 &  16 &  8 &  240 &  16+240 &  8 & 8 \\\hline
Accelerators & --- & ---&--- & ---& ---& ---& ---& --- &+K20x & +K20x  \\ \hline
\multicolumn{11}{c}{} \\
%\multicolumn{11}{c}{}\\
\end{tabular}

\begin{minipage}[t]{2.10in}\centering\par{{\bf A (MPI ranks: 64)}}\end{minipage}
\begin{minipage}[t]{2.10in}\centering\par{{\bf B (MPI ranks: 256)}}\end{minipage} 
\begin{minipage}[t]{0.80in}\centering\par{{\bf}}\end{minipage} \\
\begin{minipage}[t]{.49\columnwidth}\centering\includegraphics[width=\textwidth]{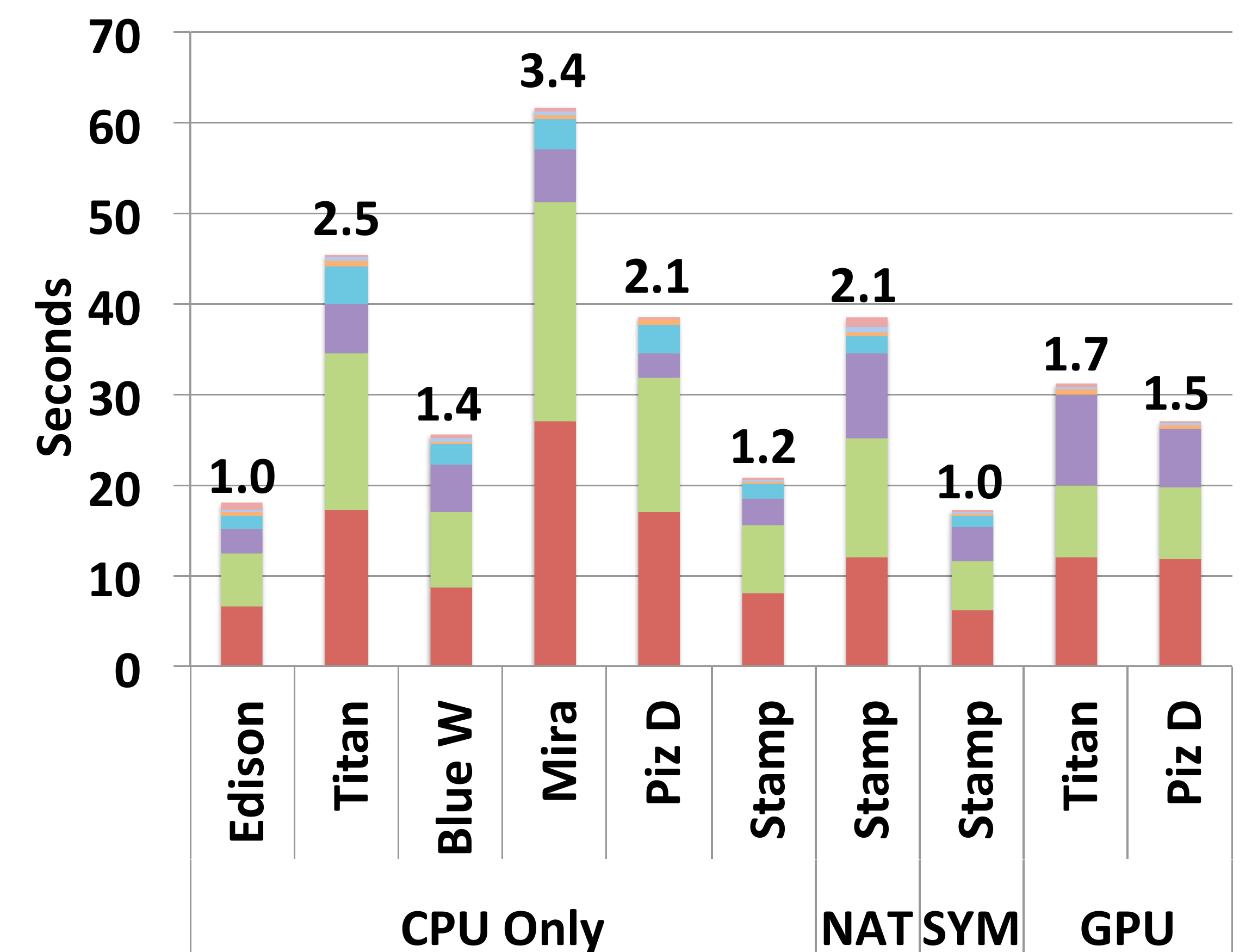}\end{minipage}
\begin{minipage}[t]{.49\columnwidth}\centering\includegraphics[width=\textwidth]{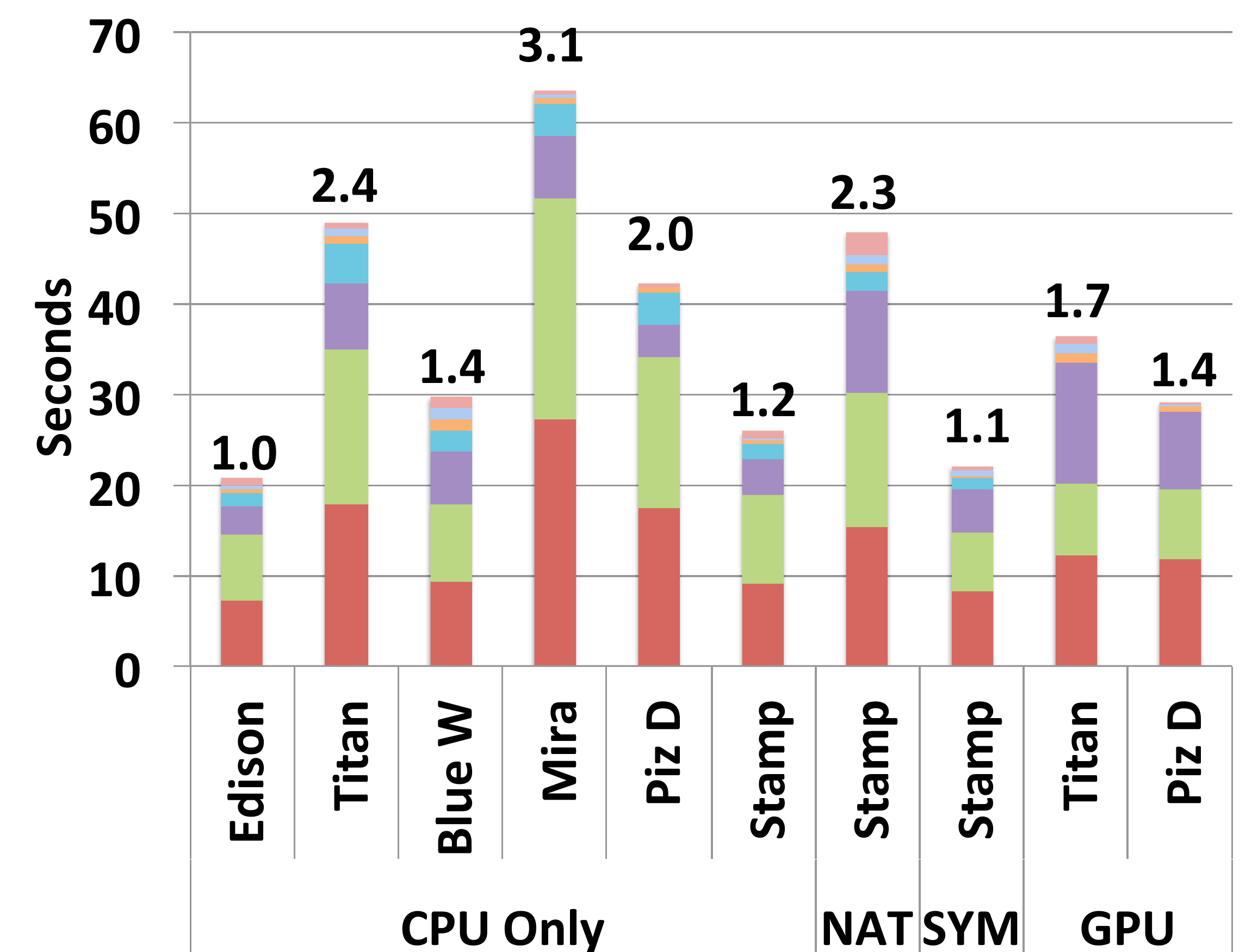}\end{minipage}
%\begin{minipage}[t]{0.80in}\centering\includegraphics[width=.99\textwidth]{../figure/legend2.pdf}\end{minipage}\\
\begin{minipage}[t]{0.40in}\centering\par{{\bf}}\end{minipage} \\
\begin{minipage}[t]{0.80in}\centering\par{{\bf}}\end{minipage} \\
\begin{minipage}[t]{2.10in}\centering\par{{\bf C (MPI ranks: 1024)}}\end{minipage}
\begin{minipage}[t]{2.10in}\centering\par{{\bf D (MPI ranks: 4096)}}\end{minipage}
\begin{minipage}[t]{0.80in}\centering\par{{\bf  }}\end{minipage} \\
\begin{minipage}[t]{.49\columnwidth}\centering\includegraphics[width=\textwidth]{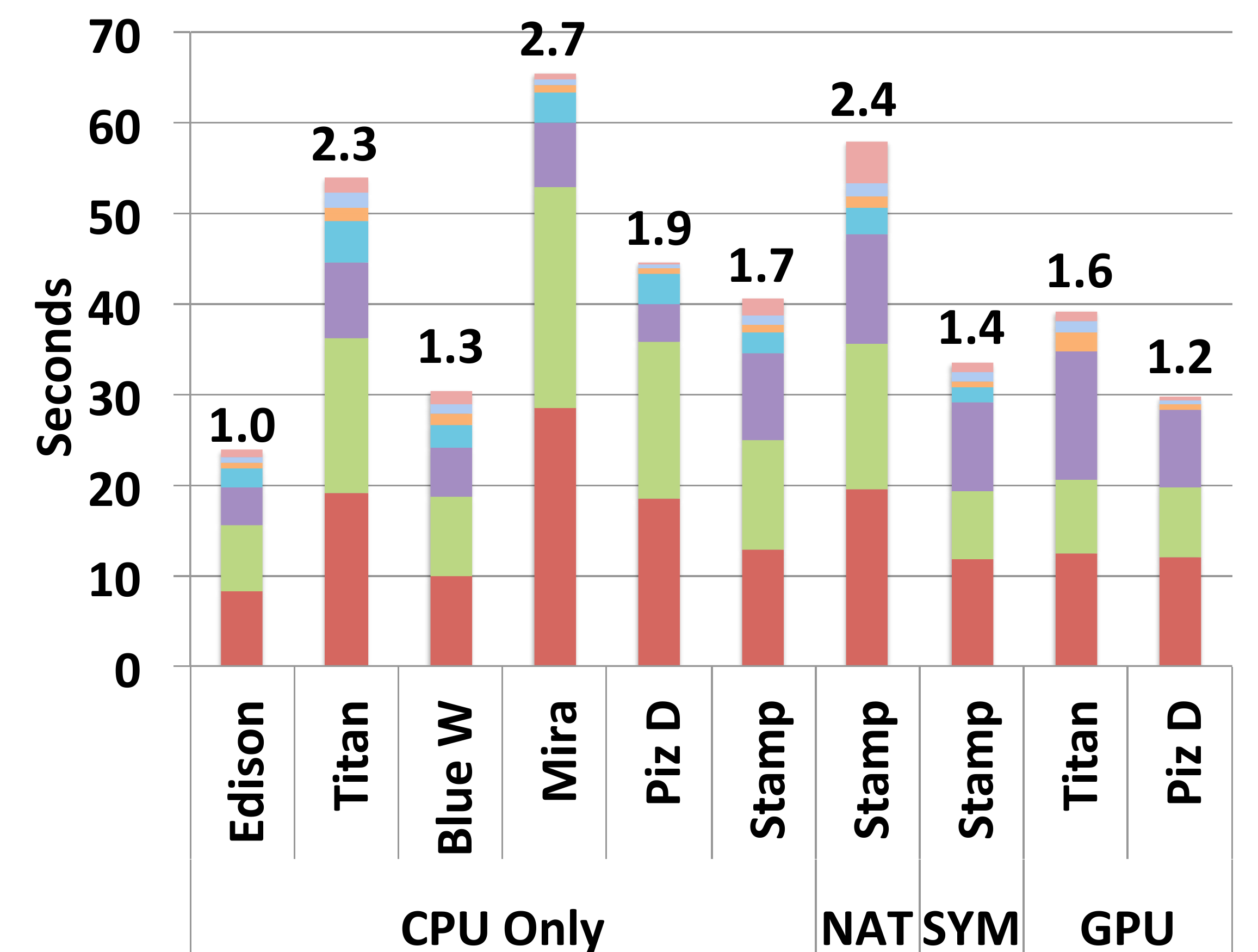}\end{minipage}
\begin{minipage}[t]{.49\columnwidth}\centering\includegraphics[width=\textwidth]{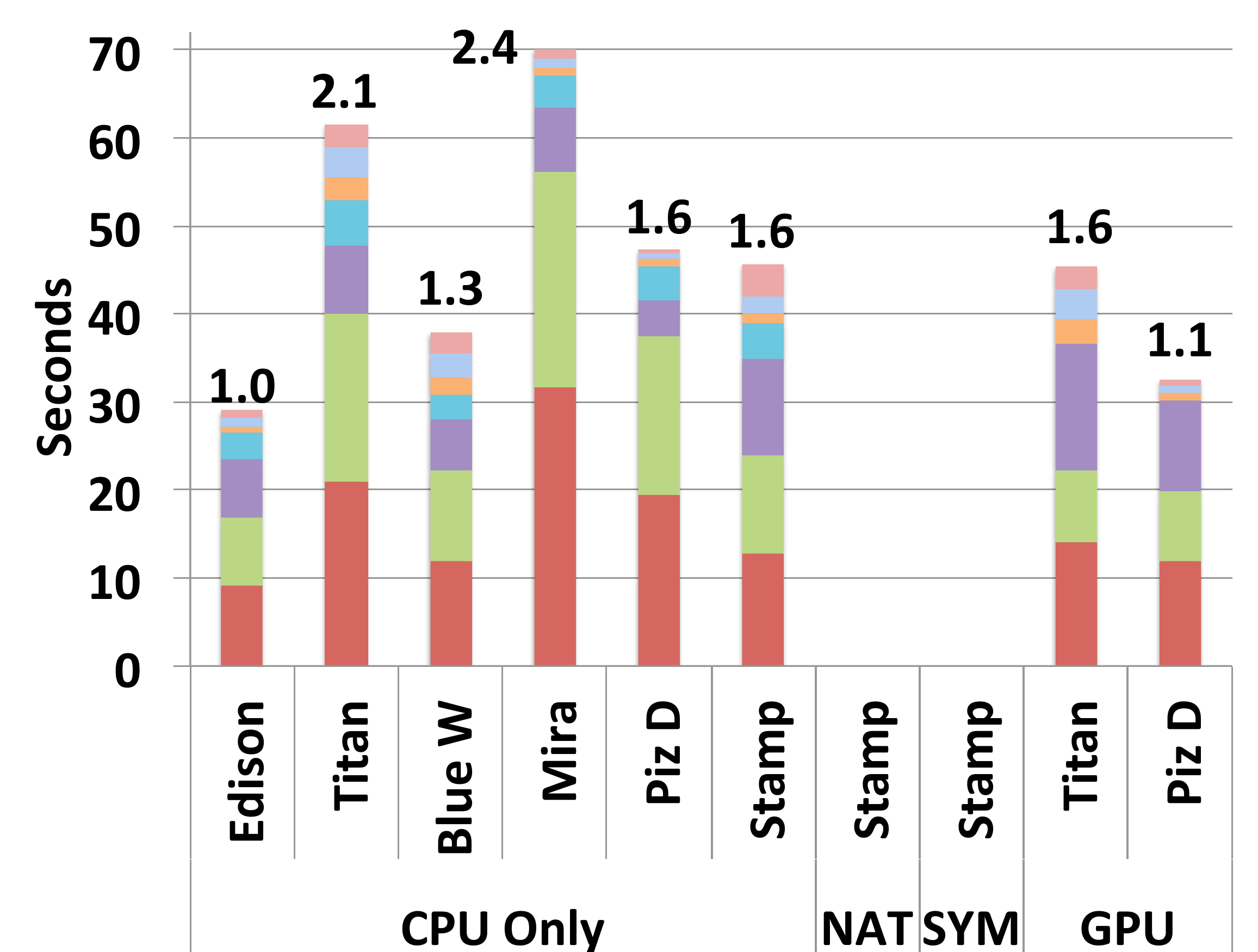}\end{minipage}
\begin{minipage}[t]{0.40in}\centering\par{{\bf}}\end{minipage} \\
\begin{minipage}[t]{3.50in}\centering\includegraphics[width=.99\textwidth]{Legend-horiz.pdf}\end{minipage}

\caption{\label{fig:weakscale2}\small
GTC-P Weak scaling run times for 100 time steps using plasma sizes A--D with 1 MPI process per compute node (2 per compute node for symmetric mode Stampede). Table presents run configurations showing MPI ranks $\times$ OpenMP threads, MIC in native and symmetric mode as well as GPU experiments. Run time relative to Edison is shown in the text above each bar (times greater than 1.0 designate a slowdown).}
\end{figure*}
}

Observe that the cache miss issue on larger size plasmas is no longer significant since the GPUs rely on massive number of threads to hide the latency. The GPU
execution times on Titan and Piz Daint are similar as both systems use the same NVIDIA K20X GPU accelerator. However, Piz Daint has shown
better PCIe performance since it has PCIe 3.0 interface whereas Titan utilizes PCIe 2.0. It worth noting that the execution time on {\bf shift}+{\bf sort} increases 2$\times$ from A to B size plasma. 
This is because A does not have radial shift where only 1-way radial partition is used. The time increase on {\bf charge} and {\bf shift} is solely related to the network performance of the systems. It shows that the Aries NIC on Piz Daint has much better
performance than the Gemini NIC on Titan since it has 3$\times$ more injection bandwidth and significantly higher global bandwidth.

\subsubsection{One Process per Node}
\label{sec:oneprocpernode}
Often, there is a desire to view a multi-socket, multi-accelerator heterogeneous compute node as the single unit of computation.
%\MPar{We should have brief discussion on overall performance ``bakeoff'' between all the systems, and more insights about the tradeoffs of the different architectural approaches : Sam included a comment on page 20 on Mira energy efficiency.}
Thus, in our second series of experiments, we use one MPI process per node (except on symmetric mode Stampede where we use 2 MPI processes per compute node). The configuration for each platform is shown at the top of Figure~\ref{fig:weakscale2}, showing MPI ranks $\times$ OpenMP threads,  MIC in native mode and symmetric mode as well as GPU experiments. These experiments help us to measure the overall node (as opposed to NUMA domain) performance, while scaling the problem to the largest node configuration possible for a given problem size.
As before, we choose 4096 MPI partitions for the D size simulation with 64-way partitions in both toroidal and poloidal dimensions. The performance results are presented 
in Figure~\ref{fig:weakscale2}, where for each problem size, the number
of computing nodes involved is the same for all systems. 
These experiments highlight the performance challenges associated with attempting to program multi-socket NUMA architectures using OpenMP.

\begin{figure}[!tb]
\centering

\begin{center}
\begin{minipage}[t]{.48\textwidth}\centering\includegraphics[width=.99\textwidth]{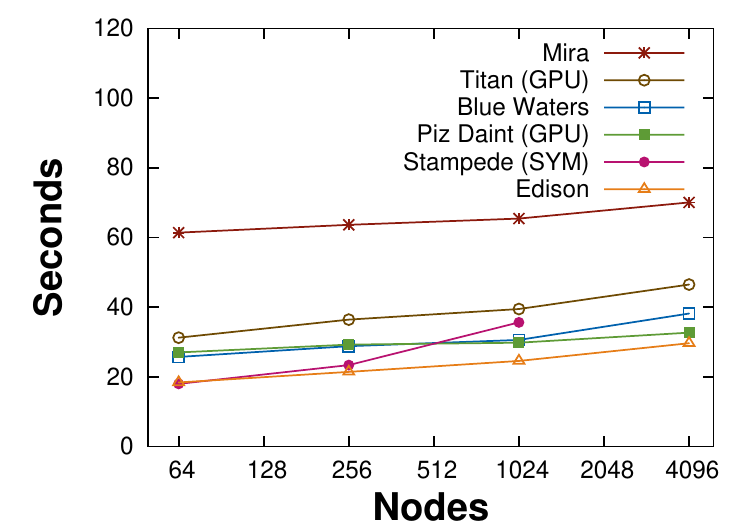}\end{minipage}
\begin{minipage}[t]{.42\textwidth}\centering\includegraphics[width=.99\textwidth]{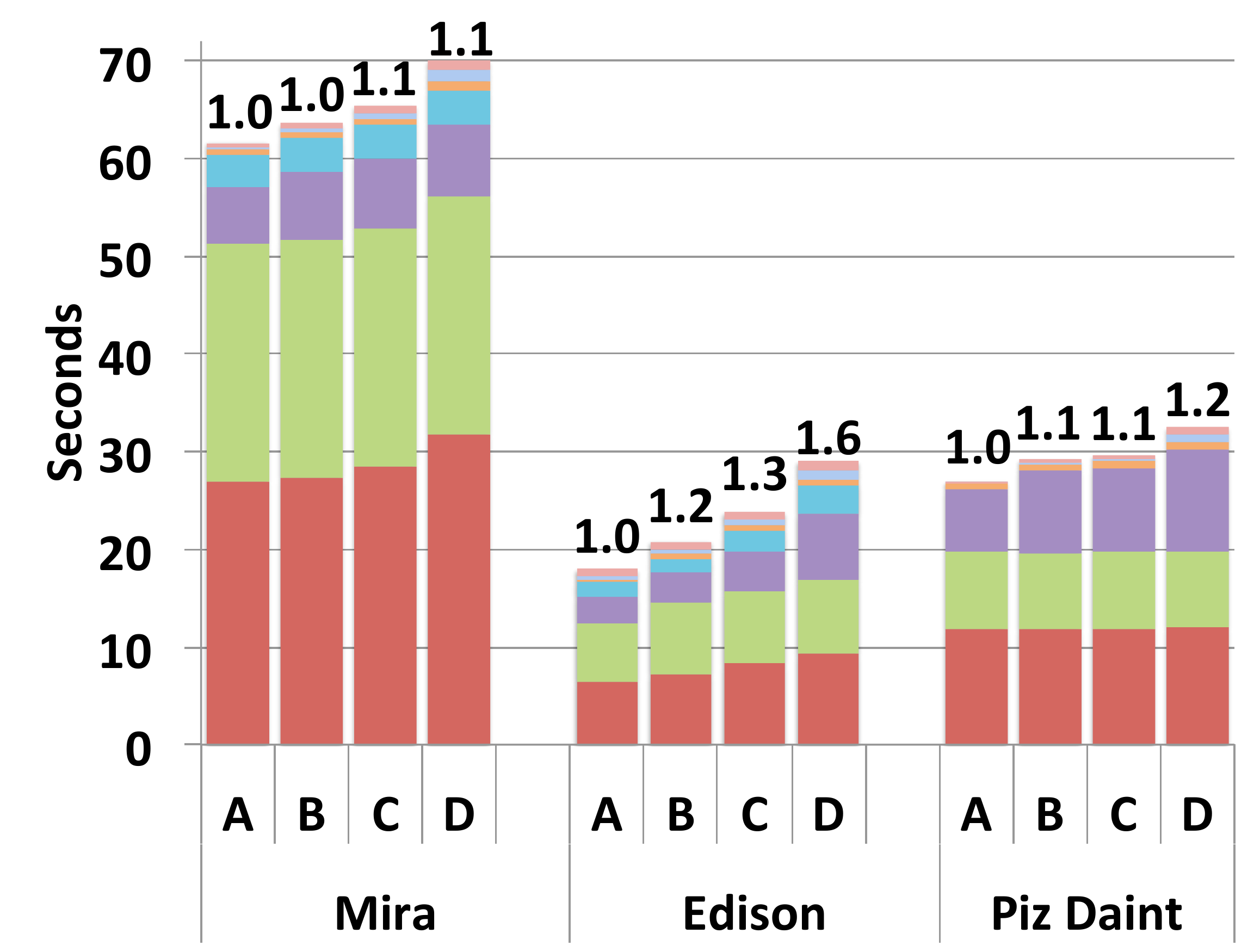}\end{minipage}
\begin{minipage}[t]{.075\textwidth}\centering\includegraphics[width=.99\textwidth]{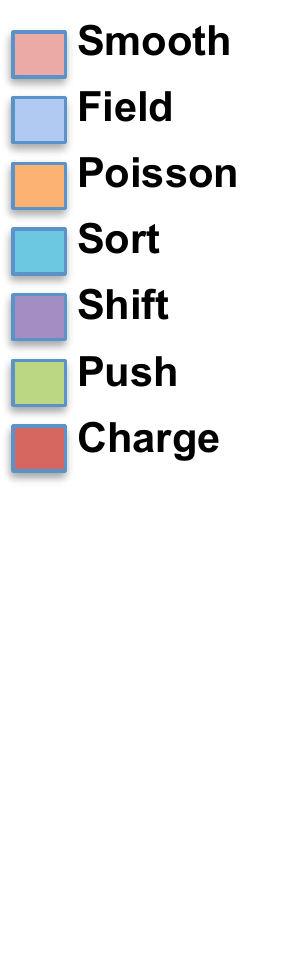}\end{minipage}

\caption{\small (left) GTC-P weak scaling run time for 100 time steps using one process per compute node. The four points on each line represent A, B, C and D size simulations from Figure~\ref{fig:weakscale2}.
 The D size data is missing for Stampede as system issues prevented us from conducting 4096 node symmetric mode experiments.  (right) Closeup of Mira (CPU), Edison (CPU), and Piz Diant (GPU) performance across A--D problem sizes (from Figure~\ref{fig:weakscale2}).   Run time relative to problem size A of each platform is shown in the text above each bar (times $>1$ designate a scaling degradation).}
\label{fig:particlescale2}
\end{center}
\end{figure}

The Xeon Phi offers three programming models --- offload (akin to GPU acceleration), native (no host), and symmetric (host and MIC are peers in a heterogeneous system).
In order to run in symmetric mode, one must balance work between CPU processes and MIC processes.
Using the data from the previous section (CPU-only and MIC-only), we observe that GTC-P performance using one MIC processor per node is comparable to the performance attained using one of the 8-core Xeon E5-2680 processors.  This suggests there is roughly a 2:1 performance difference between the Xeons and the MIC.  
When running in symmetric mode on Stampede, there are two ways of decomposing particles on a node.
%With particle decomposition for the MPI processes in the same node, there are two ways to balance the workload between the processes on the host and the MIC when running in symmetric mode. 
On one hand, we can use 1 MPI process per 8-core Xeon E5-2680 and 1 MPI process per MIC, thus resulting in 3 MPI processes per compute node in which each process has approximately the same amount of particles. On another hand, we can use 1 MPI process on two 8-core Xeon E5-2680 and 1 MPI process per Xeon Phi. 
In this section, we use the latter wherein we assign twice as many particles to the multi-socket CPU process as the MIC process.
This configuration delivered better overall performance as the grid-based subroutines are now threaded over all 16 host cores. 
Figure~\ref{fig:weakscale2} shows that the symmetric (SYM) mode boosts the performance by 1.2$\times$ with respect to the CPU-only version up to 1024 nodes on Stampede.

Figure~\ref{fig:particlescale2} presents the weak scaling trend for the simulations in Figure \ref{fig:weakscale2}, where ideal scaling is a flat line. The slope of the 
results essentially shows the impact of load imbalance (due to cache design) and network performance of a given system to our application. 
Generally speaking, we see good scalability on all systems except Stampede where there is marked degradation in scalability on its fat tree beyond 256 nodes.
Moreover, Titan and Blue Waters both use the same Gemini 3D torus.  When subtracting for the difference in the performance of 4 CPUs vs. to 2 CPUs and a GPU, we see very similar scalability between these two machines. 
Interestingly, Piz Daint and Edison show different scalability despite both using the same Cray Aries Dragonfly network.
Figure~\ref{fig:particlescale2}(right) shows there is a substantial increase in time in {\bf charge} on Edison while time remains roughly constant on Piz Daint and Mira.
An investigation of the timing showed even the average performance over 100 iterations was highly variable on Edison.
One can speculate that there is a qualitative difference in machine usage and contention, but further investigation is warranted.

Although Mira delivers the lowest performance, one should remember that it consumes roughly one quarter of the power per node as Edison and only 8\% of the full machine was utilized.
As GTC-P has demonstrated effective scalability to the full capability of Mira and Titan \cite{Wang13,Tang14}, alternate experiments in which the D100 problem is distributed across each full machine (accepting that work per node will differ from one machine to the next) is a potential avenue of future research.
%Note that both Titan and Blue Waters have Gemini 3D torus interconnect and thus we should expect similar trend. However, Figure~\ref{fig:particlescale2} shows that Blue Waters has slightly better network performance up to 1024 nodes compared with Titan. This is because we explicitly select a set of node sets with optimal network performance through job scheduler on Blue Waters \fix{and do not have this functionality on Titan?}.

\subsection{Portability vs. Speedup}
The computer architecture has evolved significantly over the last two decades. This revolution requires the application developers to modify their codes to take full advantage of the new hardware. Sometimes, this even means a complete
re-write of a code. Table \ref{tab:port} shows the trade-off between portability with respect to the number of lines changed and the achieved speedup based on our experience in porting GTC-P to different architectures.
While the GPU hardware enhances the performance by up to 4.69x, it also requires rewriting whole kernels in CUDA. Directive-based programming models, such as OpenMP, greatly simplify the implementation of new optimizations. In the case of GPU, however, the current directive models (OpenACC 2.0, OpenMP 4.0) are still lacking several key features needed to achieve the same performance as the CUDA version. Overall, the symmetric mode MIC version seems to have the best balance 
between portability and speedup. However, one has to remember that the speedup is solely obtained by utilizing additional resources.

{
\footnotesize
\setlength{\tabcolsep}{1pt}
\begin{table*}[!tb]
\begin{minipage}[b]{1.0\textwidth}\centering\footnotesize
\begin{tabular}{c|ccc|ccc|ccc}
%\hline
				& \multicolumn{3}{c|}{\bf{charge}} & \multicolumn{3}{c|}{\bf{push}} & \multicolumn{3}{c}{\bf{shift}} \\
\bf{Architecture}	& {\bf speedup} & {\bf NOLC} & {\bf \%LC} & {\bf speedup} & {\bf NOLC} & {\bf \%LC} & {\bf speedup} & {\bf NOLC} & {\bf \%LC} \\ \hline
\bf{CPU}& 1.0 & 0 & 0\% & 1.0 & 0 & 0\% & 1.0 & 0 & 0\%\\ \hline
\bf{CPU+GPU}& 1.6-2.8 & 830 & 140\% &  2.29-4.69 & 928 & 105\% & 0.78-1.55 & 555 & 150\% \\ \hline
\bf{MIC (Native)} & 1.03 & 43 & 4\% & 1.21 & 133 & 13\% & 0.56 & 69 & 6\% \\ \hline
\bf{CPU+MIC (Sym.)} & 1.47 & 43 & 4\% & 1.85 & 133 & 13\% & 1.03 & 69 & 6\% \\ \hline

\end{tabular}
\end{minipage}
\caption{Portability vs. Speedup on different compute architectures. The number of lines of code changed (NOLC) and the percentage of lines of code changed (\%LC) partially indicates the level of effort made to port and optimize GTC-P to a specific architecture. 
The speedup is obtained from the data in Figure \ref{fig:weakscale}. For example, the two GPU speedup values are calculated by comparing the kernel CPU and GPU times on Piz Daint and Titan, respectively.
The speedup can be regarded as chip-to-chip performance comparison except for the MIC symmetric mode. \label{tab:port}}
\end{table*}
}

\subsection{ITG Simulations}\label{sec:itg}
With the advances of computer architectures and numerical algorithm as well as the development
of the modern GTC-P code, we now have the ability to study long time behavior of turbulence transport on large scale devices (ITER-size) with high phase-space 
resolution. Compared with the original GTC size scaling study of 7000 time steps~\cite{Lin02}, our production run has increased the phase-space resolution by up to 12$\times$ and total time steps by 4$\times$.

\begin{figure}[tb]
\begin{center}
\includegraphics[width=.45\textwidth]{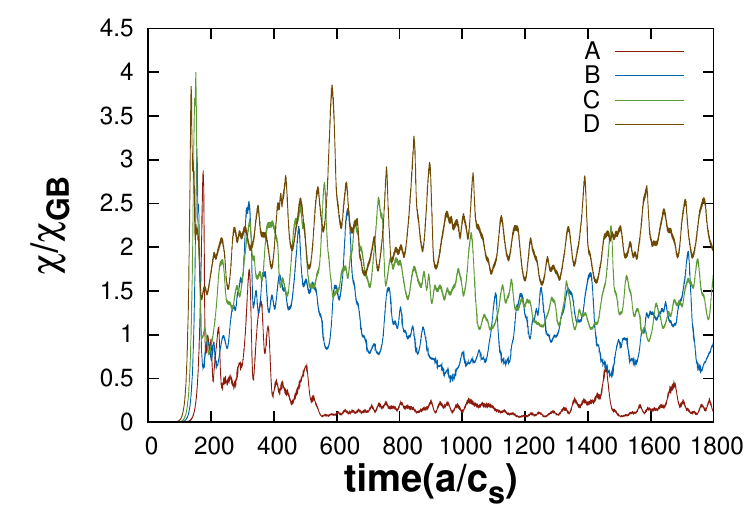}
\includegraphics[width=.45\textwidth]{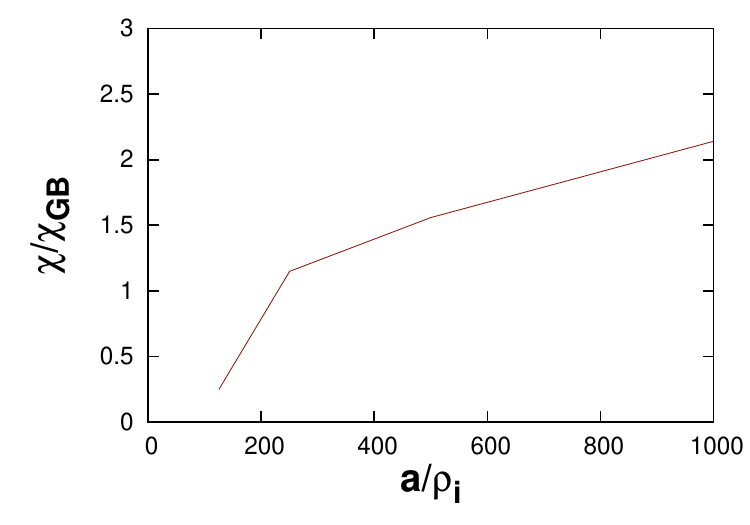}
\caption{\small\label{fig:convergence} (Left) The time evolution of the heat conductivity in Gyro-Bohm units for a wide range plasma sizes. $a$ is the minor radius of the tokamak and $c_s (= \sqrt{Te/m_i})$ is the ion acoustic speed. 
The normalized length $a/\rho_i$=125, 250, 500 and 1000 stand for A-D plasma sizes, respectively.
The normalized time 1800 corresponds to 30,000 time steps. (Right) The time average of the heat conductivity between time 240 to 1800 for different plasma sizes. 
Note this simulation was carried out on Mira using up to 32,768 compute nodes. }

\end{center}
\end{figure}

The physical parameters for the simulations are the Cyclone case~\cite{Dimits00} which has peak ion temperature gradient at $r=0.5a$ and local parameters: $R_0/L_T=6.9$,
$R_0/L_n=2.2$, $q=1.4$ and $(r/q)(dq/dr)=0.78$. Here $a$ is the minor radius, $R_0$ is the major radius, $L_T$ and $L_n$ are the temperature and density gradient 
scale length, respectively. The safe factor $q$ defines the amount of twist in the magnetic field as a function of radius. A homogeneous Dirichlet boundary condition
is enforced for the electrostatic potential at the core and edge boundary, $r=0.1a$ and $r=0.9a$. This model uses a circular cross section and assumes electrostatic
fluctuations with adiabatic electron. The pressure gradient profile, which is the driving force for the turbulence, is defined as $\exp\{-{[r-0.5a]/0.35a]}^6\}$. The 
velocity space nonlinearity, which is usually ignored as a high order term, but has been shown to be important to maintain energy conservation~\cite{Dimits00}, is included in the 
simulations. 
%Different from the model in GTC, the simulations employ remapping technique \cite{Chen07,Wang11} to stabilize long time simulations. A more in depth study of the effects of various dissipation models \cite{Chen07,Wang11,Mcmillan08} will be presented elsewhere.
Figure \ref{fig:convergence} shows the time evolution of the diffusivity in Gyro-Bohm units for a wide range plasma sizes up to 30,000 time steps. We see a gradual \textquotedblleft rollover\textquotedblright\ of transport level in large size devices. 
However, this \textquotedblleft rollover\textquotedblright\ is much more gradual and the transport level's magnitude is significant lower than the value obtained 
from much lower resolution and shorter-duration kinetic simulation by the original GTC code early \cite{Lin02}. It is unclear at this point what are the exact reasons that cause the transport level difference between
these two codes. Considering that the original GTC used a heat bath model whereas GTC-P used numerical dissipation
 to stabilize the long time simulation, we are systematically investigating the effects of various dissipation models. A more in depth study will be presented elsewhere.

\section{Conclusion}
In this paper, we have described our efforts in developing a highly parallelized PIC code which solves the 5D Vlasov-Poisson equation with efficient utilization of modern massively parallel HPC systems.
In particular, we discuss the kernel implementation on NVIDIA GPU accelerators and Intel Xeon Phi co-processors. 
As being \textquotedblleft memory bound\textquotedblright,~the performance of our application is correlated to the DRAM STREAM bandwidth per NUMA node on CPU. On accelerators, however, its performance is substantially lower than the STREAM-predicted number due to data locality, synchronization and PCIe challenges in 
\textquotedblleft gather\textquotedblright~and \textquotedblleft scatter\textquotedblright~operations and unavoidable data transfer through PCIe.
Comparisons of the code performance on multiple supercomputer systems show that the network performance of a system 
has become an increasingly important factor for the overall performance of our application, especially for large scale simulations. 
GTC-P currently uses two-sided MPI for distributed memory communication. In near-future work, we will actively explore the potential benefits of deploying one-sided communication using MPI standard 3.0.
Our future work will also involve the development of the offload version for Intel Xeon Phi based system using both Intel compiler offload directives and OpenMP 4.0 standard. Additionally, we will explore the utilization of multiple Intel Xeon Phi co-processors in a single node. 
Our expectation is that if these efforts prove successful, scientific discovery in application domains such as fusion energy will be significantly enhanced.

\section*{Acknowledgments}
The authors would like to thank Rezaur Rahman from Intel Corporation for developing the low-level intrinsics on Intel Xeon Phi (MIC) for the GTC-P code.  
Dr. Wang and Dr. Tang were supported by the NSF OCI-1128080/G8 Initiative: G8 Research Councils Initiative on Multilateral Research Funding. 
Contributions from Dr. Ethier and Dr. Tang (in part) were supported by DOE contract DE-AC02-09CH11466 at the Princeton Plasma Physics Laboratory (PPPL). 
Co-authors from the Lawrence Berkeley National Laboratory (LBNL) were supported by the DOE-SC funds from  contract number DE-AC02-05CH11231. 
The authors would also like to extend their gratitude to the modern supercomputing facilities at (i) the Swiss National Supercomputing Center for access 
to \textquotedblleft Piz Daint\textquotedblright; (ii) NSF's Texas Advanced Computing Center (TACC)) for access to \textquotedblleft Stampede\textquotedblright;
 and (iii) NSF's National Center for Supercomputing Applications (NCSA) for access to \textquotedblleft Blue Waters\textquotedblright. 
This research benefited greatly from a significant award of computer time at the Argonne Leadership Computing Facility (ALCF) provided by the DOE-SC Innovative and 
Novel Computational Impact on Theory and Experiment (INCITE) program.  Resources provided by the national user facility at the Lawrence Berkeley National 
Laboratory's National Energy Research Scientific Computing Center (NERSC) and by the Oak Ridge Leadership Computing Facility (OLCF) are also gratefully acknowledged.
\begin{singlespace}
\bibliographystyle{abbrv}
\bibliography{gtc13_IJHPCA}

\end{singlespace}

\end{document}